\documentclass[namedreferences,hyperref,optionalrh]{spr-sola}
\usepackage{amsmath}
\usepackage{graphicx}        
\usepackage{color}           

\hypersetup{
colorlinks,
citecolor=blue,
filecolor=blue,
linkcolor=blue,
urlcolor=blue
}

\chardef\us=`\_

\begin{document}

\begin{frontmatter}
\title{Modeling Decadal and Centennial Solar UV Irradiance Changes}

\author[addressref=aff1, email={raffaele.reda@roma2.infn.it}]{\inits{R.}\fnm{Raffaele}~\snm{Reda}\orcid{0000-0001-8623-5318}}
\author[addressref={aff1}, email={penza@roma2.infn.it}]{\inits{V.}\fnm{Valentina}~\snm{Penza}\orcid{0000-0002-3948-2268}}
\author[addressref=aff2, email={scriscuo@nso.edu}]{\inits{S.}\fnm{Serena}~\snm{Criscuoli}\orcid{0000-0002-4525-9038}}
\author[addressref=aff2, email={luca@nso.edu}]
{\inits{L.}\fnm{Luca}~\snm{Bertello}\orcid{0000-0002-1155-7141}}
\author[addressref=aff1, email={matteo.cantoresi@roma2.infn.it}]
{\inits{M.}\fnm{Matteo}~\snm{Cantoresi}\orcid{0000-0003-4898-2683}}
\author[addressref=aff1, email={lorenza.lucaferri@roma2.infn.it}]
{\inits{L.}\fnm{Lorenza}~\snm{Lucaferri}\orcid{0009-0000-9757-8455}}
\author[addressref=aff3, email={ulzg@zhaw.ch}]
{\inits{S.}\fnm{Simone}~\snm{Ulzega}\orcid{0000-0002-8744-4266}}

\author[corref, addressref=aff1,email={berrilli@roma2.infn.it}]{\inits{F.}\fnm{Francesco}~\snm{Berrilli}\orcid{0000-0002-2276-3733}}

\address[id=aff1]{Dipartimento di Fisica, Universit\`a degli Studi di Roma Tor Vergata, Via della Ricerca Scientifica 1, Roma, 00133, Italy}
\address[id=aff2]{National Solar Observatory, 3665 Discovery Dr., Boulder, CO 80303, USA}
\address[id=aff3]{Institute of Computational Life Sciences, Zurich University of Applied Sciences (ZHAW), 8820 Wädenswil, Switzerland}

\runningauthor{Reda et al.}
\runningtitle{UV solar Bands}

\begin{abstract}
Reconstructions of solar spectral irradiance—especially in the ultraviolet (UV) range—are crucial for understanding Earth’s climate system. 
Although total solar irradiance (TSI) has been thoroughly investigated, the spectral composition of solar radiation offers a deeper insight into its interactions with the atmosphere, biosphere, and climate. UV radiation, in particular, plays a key role in stratospheric chemistry and the dynamics of stratospheric ozone. Reconstructing solar irradiance over the past centuries requires accounting for both the cyclic modulation of active-region coverage associated with the 11-year solar cycle and the longer-term secular trends, including their centennial variability.
This study utilizes an empirical framework, based on a 1000-year record of Open Solar Flux, to characterize the various temporal components of solar irradiance variability. We then combine these components to reconstruct Solar UV irradiance variations in spectral bands crucial for Earth's atmospheric studies.
\end{abstract}

\end{frontmatter}

\section{Introduction}
     \label{S-Introduction}

Solar irradiance constitutes the principal energy source for the Earth system, engaging in complex interactions with the terrestrial atmosphere and exerting an influence on climate over long timescales. 
Extensive efforts have been dedicated to studying Total Solar Irradiance (TSI)—the Sun’s irradiance integrated across the entire wavelength spectrum—using both observations and models.
However, specific spectral bands — particularly in the ultraviolet (UV) — are critically important across a range of disciplines, including solar and stellar physics, atmospheric modeling, climatology, and the study of habitability conditions on exoplanets \citep[see e.g.,][]{Pienitz2000, linsky2014, lovric2017, Galuzzo2021, Reda2023}.

The TSI exhibits typical variations of about 0.1\% from minimum to maximum of the 11-year cycle \citep[e.g.][]{kopp2016}, whereas variations in more energetic spectral bands, such as the UV, can be significantly larger, reaching up to 10\% \citep{frohlich2004, criscuoli2019, Woods21}. These variations are particularly important because of the strong interaction of solar radiation at these wavelengths with Earth's upper and middle atmosphere. According to conventional nomenclature, it is possible to divide the UV band into several regions: middle UV (MUV, 180–300 nm), far UV (FUV, 115–180 nm) and extreme UV (EUV, 10–115 nm).
Slightly different definitions of these bands, particularly MUV and FUV, can be found in literature. In this work, we use the definitions adopted by the SORCE-SOLSTICE experiment, for which MUV and FUV data are available \citep{McClintock2005}\footnote{\url{https://lasp.colorado.edu/sorce/files/2020/04/SORCE_SOLSTICE_Release_Notes_for_Version_17.pdf}}.

In particular, EUV radiation can affect ionospheric and thermospheric structures, and thereby influence, for example, the propagation of radio signals \citep[e.g.][]{Bigazzi2020}.
UV radiation also plays a key role in atmospheric density modeling and forecasting, directly influencing satellite orbits due to increased atmospheric drag.
In the case of the stratosphere, radiation in the FUV and MUV bands can change ozone levels. 
In particular, ozone formation is primarily driven by high-energy UV photons with wavelengths shorter than 240 nm, falling within the Schumann-Runge bands and Herzberg continuum. These photons dissociate molecular oxygen $O_{2}$, enabling the subsequent creation of ozone $O_{3}$ through the reaction between $O_{2}$ and atomic oxygen $O$.
Conversely, ozone molecules can be broken apart by lower-energy UV radiation, particularly at wavelengths up to 320 nm (Hartley-Huggins bands). However, this is not the main pathway by which stratospheric ozone is destroyed, because $O_{2}$ can easily recombine with $O$. Additional chemical processes acting at different altitudes - potentially involving catalytic reaction cycles - take part in the radiative interplay that governs the vertical distribution of ozone and its role in stratospheric heating and chemical feedbacks \citep{chapman1930, Haigh2007}. Due to the aforementioned mechanisms, the greater variability of the solar cycle in the UV part of the spectrum has a stronger modulating effect on ozone production than on its destruction. This results in a 2–4\% increase in ozone abundance in the upper stratosphere during periods of solar maximum \citep{lilensten2015}.\\ The FUV and MUV bands also enable the calculation of the spectral color index FUV-MUV. This descriptor, introduced by \cite{lovric2017} and further investigated by \cite{Criscuoli2018}, can be used to characterize stellar UV emission and shows a strong correlation with the Mg~II index. \\
This underscores the importance of direct observations for studying solar UV variations. However, systematic measurements of solar UV radiation have only been performed since the advent of space-based telescopes in the late 1960s, with missions such as OSO-1 \citep{Neupert1965}, NIMBUS-7 \citep{Cebula1988}, SORCE \citep{McClintock2005} and TSIS \citep{Richard2024}. Therefore, reconstructions based on solar and terrestrial proxies are particularly valuable for investigating solar variability in the past centuries \citep{kakuwa}. \\
It is commonly accepted that Total Solar Irradiance, and Spectral Solar Irradiance in the visible and near-infrared variations on short to medium timescales, from days to decades, are primarily driven by the presence of dark (sunspots) and bright (network and plage) regions on the solar surface, modulated by the 11-year solar cycle \citep[e.g.][]{Steinegger1996,Berrilli1999,Ermolli2003, Criscuoli2023, marchenko2024}. On longer, secular timescales, variations are also evident, as demonstrated by grand minima and maxima observed in sunspot records and magnetic activity proxies such as radionuclides \citep[e.g.,][]{Stuiver1998, Vonmoos2006, usoskin2016, usoskin2017, Vecchio2017, Petrie2021}. Modeling these variations provides a key input for understanding the global dynamics of the Earth’s climate system \citep[see e.g.,][]{lockwood2012, solanki2013,Bordi2015,matthes2017,liu2023}. \\
In this paper, we reconstruct four UV spectral bands (100-243 nm, 243-308 nm, FUV and MUV) over the past millennium using the same approach adopted in \cite{penza2022} and \cite{penza2024}, where the method was applied to reconstruct the TSI. 
The applicability of this approach to spectral UV irradiance, beyond its original use for TSI reconstruction, is supported by several studies \citep[e.g.,][]{krivova2003, wenzler, ball2011, fontenla2011, lean2020}. These suggest that solar irradiance variability across all wavelengths is entirely driven by the evolution of the magnetic field at the solar surface, which can be represented by the combined contributions of a limited set of magnetic structures, such as sunspots, faculae, and the network. 
Moreover, in the UV range, it is well established that irradiance variability is predominantly driven by bright features such as faculae and network regions, while the contribution from sunspots is almost negligible. This approach has also enabled the reconstruction of specific UV proxies, such as the Mg II index \citep[e.g.][]{Lean1997,Criscuoli2018, berrilli2020,Sowmya2025}.

Specifically, we employ the same composite of active region coverages \citep{chatzistergos20, mandal2020} and the same composite of Open Solar Flux $F_{0}$, with the latter serving as a proxy for tracking long-term changes on time scales larger than the 22-year solar magnetic cycle.
The composite of $F_{0}$ for the period 971–1899 C.E. is derived from the reconstruction by \citet{usoskin2021}, which is based specifically on the cosmogenic isotope $^{14}\mathrm{C}$. The extension to the present (1900–2020 C.E.) is obtained using an empirical relation between $F_{0}$ and the solar modulation potential $\phi$, given by \citet{muscheler2007} for 1513–1949 C.E. and by \citet{usoskin2017} for 1950–2025 C.E.. The reconstruction of $F_{0}$ for the post-1900 period has been validated by comparison with observational data from \citet{owens2017}.

\section{Reconstruction of UV Bands Over the Solar Cycle } 
      \label{rec_1}      

\subsection{UV Data and Selected Bands} 
  \label{data}

The UV data used in this work to parameterize the reconstruction are from the SSI3 composite by \citet{Woods21}. We refer to the website of the Laboratory for Atmospheric and Space Physics (LASP) \footnote{\url{https://lasp.colorado.edu/lisird/data/lasp_gsfc_composite_ssi}} for a detailed description of the instrumentation and the composite construction.

\begin{figure}[h]    
\centerline{\includegraphics[width=\textwidth]{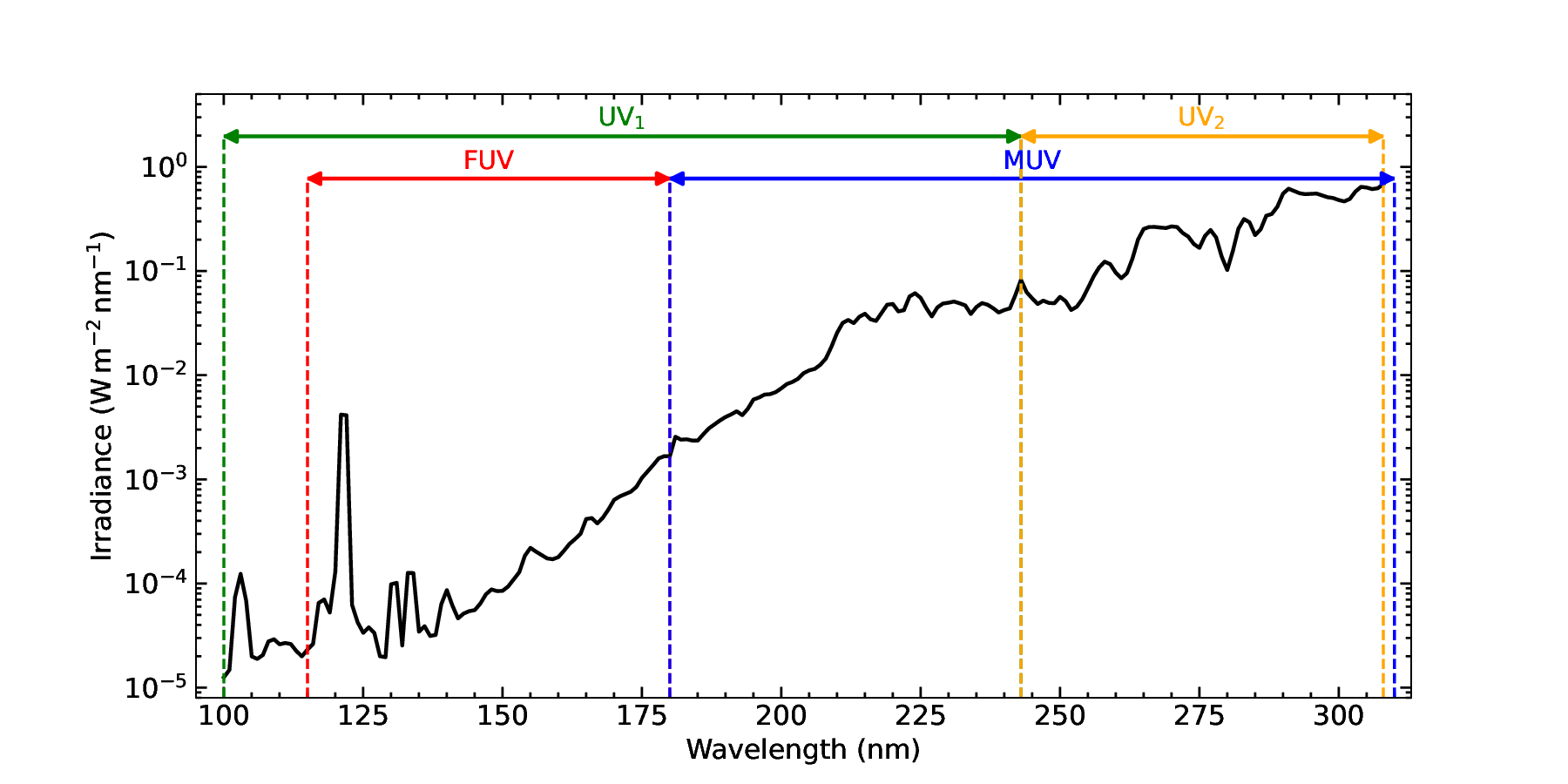}}

\small
        \caption{An example of a SSI3 composite spectrum, in logarithmic scale, from 100 nm to 310 nm with different bands highlighted by different colors.}
\label{UV_sorce}
\end{figure}

\begin{figure}[h]    
\centerline{\includegraphics[width=0.7\textwidth]{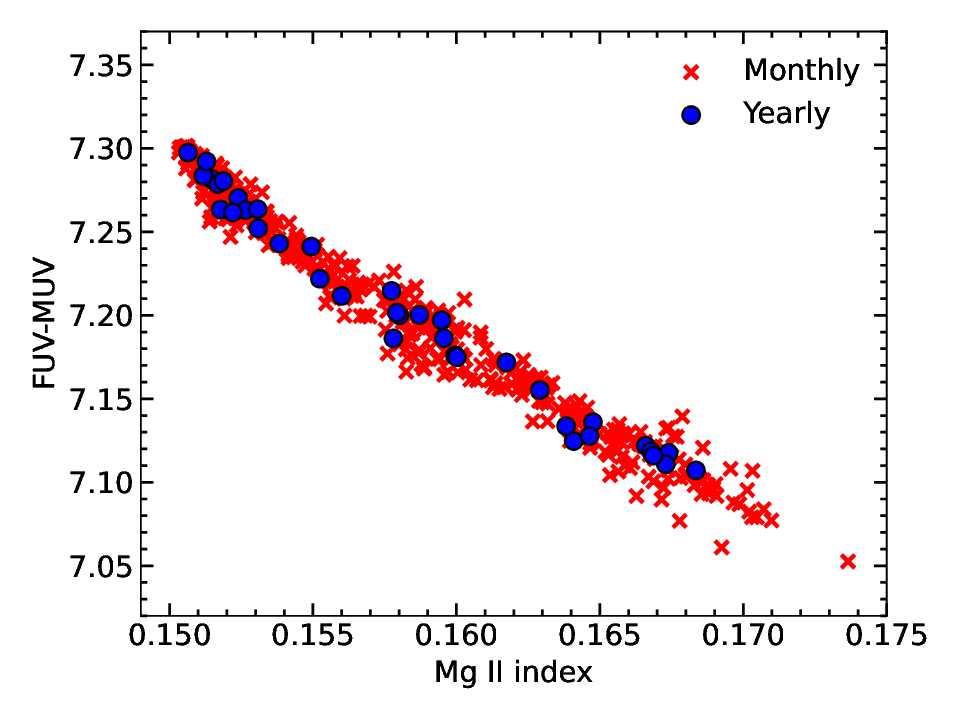}}
\small
        \caption{The correlation between FUV-MUV color and the Bremen Mg II index is shown for monthly values (red crosses) and for yearly averaged data (blue points) for the period 1980-2016. The Pearson's coefficient is r=-0.985 for monthly values and r=-0.989 for yearly values.}
\label{UV_sorce_corr}
\end{figure}

In particular, we consider UV data for the period 1980–2020 C.E. and in the wavelength range 100–310~nm, within which we identify four specific bands of interest:

\begin{itemize}
  \item FUV (Far Ultraviolet): 115--180~nm;
  \item MUV (Middle Ultraviolet): 180--310~nm;
  \item $UV_{1}$: 100--243~nm;
  \item $UV_{2}$: 243--308~nm.
\end{itemize}

An example of an SSI3 composite spectrum is shown in Fig.~\ref{UV_sorce}, where the four bands mentioned above are highlighted using different colors. \\
The FUV and MUV bands allow the calculation of the spectral color index, defined as:

\begin{equation}  
\label{FUV_MUV}
 \mathrm{FUV-MUV} = -2.5\; \mathrm{log} \frac{F_{fuv}}{F_{muv}} + Z_{fuv} + Z_{muv}
\end{equation}

where $F_{fuv}$ and $F_{muv}$ are the fluxes integrated in the corresponding bands. In this work, the zero-point magnitudes, $Z_{fuv}$ and $Z_{muv}$, are arbitrarily set to zero. This applies a constant vertical shift to all magnitude values but does not affect the subsequent results.

This quantity can be used to characterize the UV emission and is strongly correlated with the Mg~II index \citep{lovric2017, Criscuoli2018}. The $UV_{1}$ and $UV_{2}$ bands are selected due to their role in the atmospheric ozone cycle: the radiation in the $UV_{1}$ range is primarily responsible for the photodissociation of molecular oxygen ($O_2$), initiating ozone ($O_3$) production, while $UV_{2}$ radiation contributes, together with $UV_{1}$, to both ozone formation and destruction through photodissociation of ozone itself. \\
The correlation of the FUV-MUV color index with the Mg II Bremen composite index\footnote{\url{https://lasp.colorado.edu/lisird/data/bremen_composite_mgii}} is shown in Fig.~\ref{UV_sorce_corr}, where a strong anti-correlation is clearly visible, with a Pearson correlation coefficient of $r = -0.989$. The best-fit linear regression of the correlation is in the form
\begin{equation}
   \mathrm{ FUV-MUV = \alpha\;Mg\,II + \beta }
\end{equation}
where in the case of monthly values $\alpha=-10.63 \pm 0.09$ and $\beta=8.89 \pm 0.01$, while for yearly values $\alpha=-10.6 \pm 0.3$ and $\beta=8.88 \pm 0.04$. \\


\subsection{Reconstruction of UV Bands Along the Last 4 Cycles} 
      \label{rec1_1}

Solar irradiance is reconstructed as the sum of contributions from various solar surface components, each weighted by its respective coverage area:
\begin{equation}
\label{Flux_rec}
F(\lambda, t) =  \sum_{j} \alpha_{j}(t) F_{j}({\lambda}),
\end{equation}

where $F_{j}(\lambda)$ is the flux at wavelength $\lambda$ from the $j^{\mathrm{th}}$ feature (assumed time-independent), and $\alpha_{j}(t)$ is the corresponding coverage.  The number and type of these components depend on the spectral band under consideration.
Given that UV irradiance variability is primarily dominated by bright features, we can expand Eq. \ref{Flux_rec} as follows:

\begin{equation}
\label{Flux_UV}
F(\lambda_{UV},t) = \alpha_{f}(t) F_{f}(\lambda_{UV}) +  \alpha_{n}(t) F_{n}(\lambda_{UV}) + (1 - \alpha_{f}(t) - \alpha_{n}(t)) F_{q}(\lambda_{UV})
\end{equation}

where the subscripts $f$, $n$ and $q$ indicate facular, network and solar quiet contributions, respectively. Then, the relative variation of the flux at a given wavelength has a very simple form:
\begin{equation}
\label{delta_Flux_UV}
\delta F(\lambda_{UV},t) \equiv \frac{F(\lambda_{UV},t) -   F_{q}(\lambda_{UV})}{F_{q}(\lambda_{UV})} = \alpha_{f}(t) \delta^{UV}_{f} +  \alpha_{n}(t) \delta^{UV}_{n}
\end{equation}
where $\delta^{UV}_{f}$ and $\delta^{UV}_{n}$ are the contrasts relative to the quiet Sun in the UV band of interest. \\
As shown in \cite{penza2022} and \cite{penza2024}, we adopt a linear relation between $\alpha_{f}(t)$ and $\alpha_{n}(t)$ \citep[e.g.][]{Criscuoli2016,Criscuoli2018,pooja2021,ermolli2022}:
\begin{equation}
 \label{an-af}
 \alpha_{n}(t) = A_{n} +  B_{n} \alpha_{f}(t).
\end{equation}
This relationship should be regarded as a simplification of the actual and complex correlation between magnetic structures. Anyway, the existence of a clear correlation between network and plage is well illustrated in Fig.~13 of \citet{ermolli2022}. Moreover, Fig.~4 of  \citet{Criscuoli2016} shows that pixels with different magnetic field strengths scale linearly with the sunspot number, implying that pixels with different magnetic field strengths should also scale linearly with each other. That allows us to rewrite Eq. \ref{delta_Flux_UV} as:
\begin{equation}
\label{delta_Flux_UV2}
\delta F(\lambda_{UV},t) = \alpha_{f}(t) \delta^{UV}_{fn} +  C^{UV}_{n}
\end{equation}
where $\delta_{fn} \equiv \delta^{UV}_{f} + B_{n} \delta^{UV}_{n}$ is the mixed facular and network contribution, while $C^{UV}_{n} \equiv A_{n} \delta^{UV}_{n}$ represents the value of the network contrast weighted by the corresponding fractional area at the solar minimum.
As in \cite{penza2024}, we use the plage coverage composite provided by \cite{chatzistergos20}\footnote{available at \url{https://www2.mps.mpg.de/projects/sun-climate/data.html}} for this reconstruction. In order to obtain an estimation of the parameters $\delta_{fn}$ and $C^{UV}_{n}$, SSI3 observations of cycle 22 and 23 were separately fit using Eq. 7. The mean of two best-fits gives the values reported in Table \ref{tab1}.

\begin{table}
\caption{Values of the parameters of Eq.\ref{delta_Flux_UV2}, obtained by best fitting to SSI3 data separately for solar cycles 22 and 23.}
\label{tab1}
\begin{tabular}{ccc}     
                         
\hline                     
Band   &   $\delta_{fn}$ & $C_{n}$ \\
 \hline
FUV & 4.82 $\pm$ 0.05  & 0.005 $\pm$ 0.001  \\
MUV &  0.17 $\pm$ 0.1  &   0.004 $\pm$ 0.001 \\
$UV_{1}$ &   0.64 $\pm$ 0.01 & 0.004 $\pm$ 0.001 \\
$UV_{2}$ & 0.17 $\pm$ 0.01  & 0.004 $\pm$ 0.001 \\
\hline
\end{tabular}
\end{table}

\begin{figure}[h]
\centering
\includegraphics[width=\textwidth]{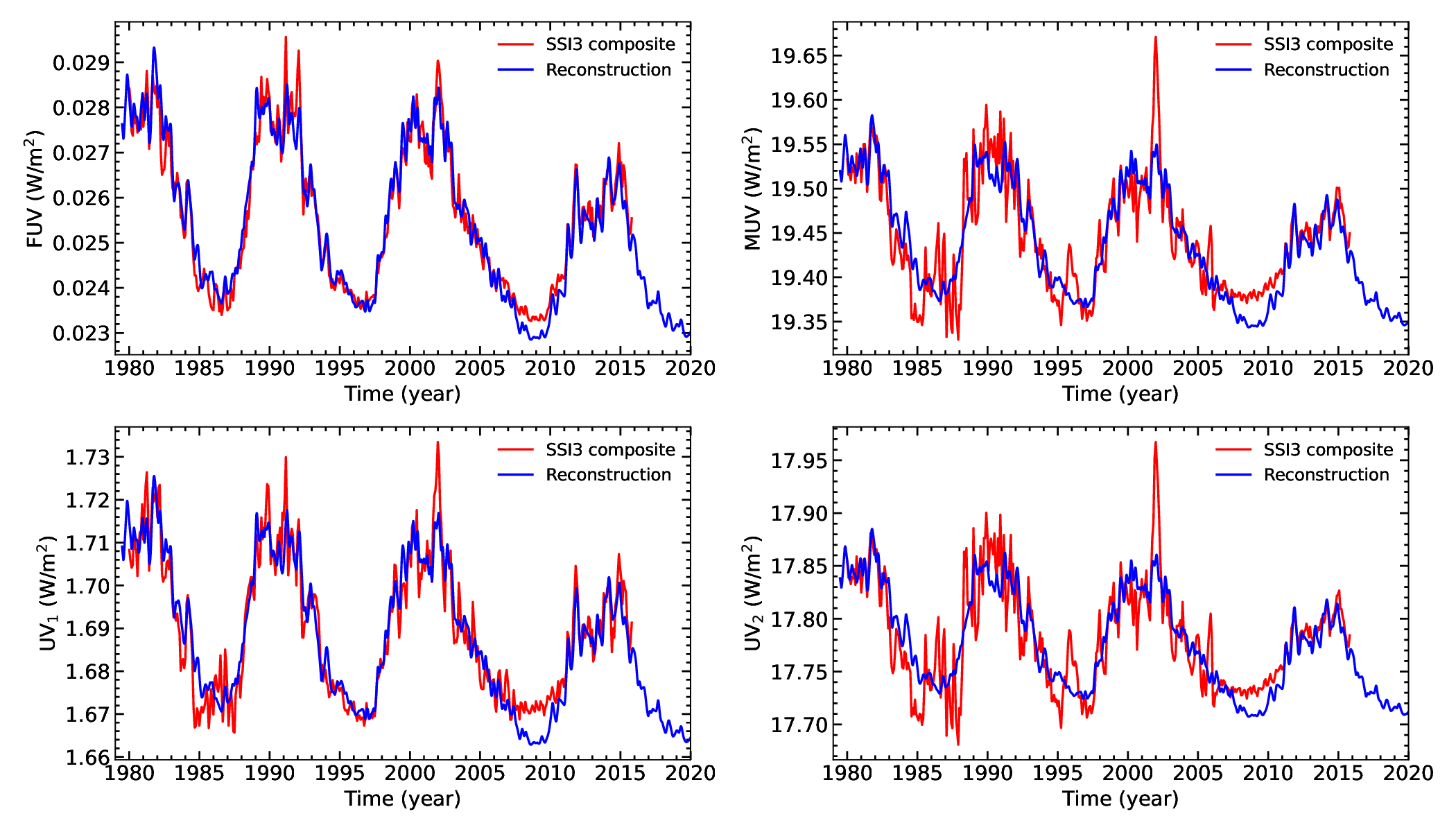}
\caption{Comparison of the monthly averages of the reconstruction (blue line) and the SSI3 composite data (red line) for the four UV bands: FUV (top-left), MUV (top-right), $\mathrm{UV_{1}}$ (bottom-left), and $\mathrm{UV_{2}}$ (bottom-right).}
\label{UV_1979}
\end{figure}

Although we have imposed no constraints on the model, these values are compatible with those reported in literature \citep{Lean1997b, fontenla1999}.
We show the different reconstructions, obtained with the approach described above, in Fig.~\ref{UV_1979}, where we compare them with the corresponding SSI3 composite data. The figure highlights that the longer-wavelength data exhibit an anomalous increase in irradiance around 2002 (the second peak of Solar Cycle 23), a feature absent at shorter wavelengths (FUV). We suppose that this behavior is not physical, but rather the result of an instrumentation artifact. However, we chose to retain these outlier data, considering that their short duration does not affect the integrity of our reconstruction. \\

\section{Reconstruction of UV Bands Along the Last Millenium} 
      \label{rec_1000}      

\subsection{Plage Coverage R
econstruction} 
  \label{plage_rec}
To reconstruct the UV spectral bands over the past millennium, we use the plage coverage time series previously reconstructed by \cite{penza2024}. In the following, we summarize the key aspects of that reconstruction.

\begin{figure}[h]
\centering
\includegraphics[width=\textwidth]{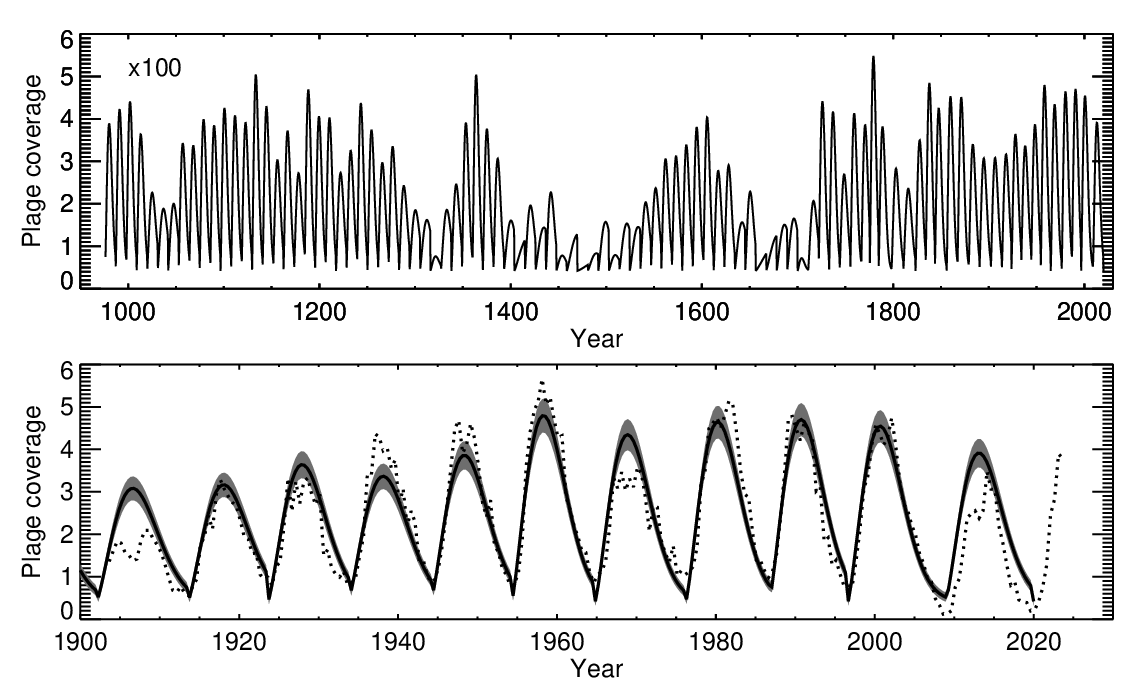}
\caption{Top panel: Reconstruction of plage area coverage from 971 CE to the present. Bottom Panel: The reconstructed
time series (solid curve) and its uncertainty range (grey region) are shown from 1900 to the present. The actual measured
monthly plage area coverage by \cite{chatzistergos20} is shown as dotted line.
}
\label{plage}
\end{figure}

\begin{enumerate}
  \item \textit{Definition of a Parametric Functional Form of the Cycle} \\

We reproduce sunspot coverage of the k-cycle by using the functional form suggested by \cite{Volobuev}:
  \begin{equation}
  \label{cycle_form}
  x_{k}(t) =  \left(\frac{t - T0_{k}}{Ts_{k}}\right)^{2} \textrm{exp}
  \left[-\left(\frac{t - T0_{k}}{Td_{k}}\right)^{2} \right]
  \quad T0_{k} < t < T0_{k} + \tau_{k}
  \end{equation}
In reality, there is a relationship between the two free parameters $Ts_{k}$ and $Td_{k}$. This relation reflects the well-known Waldmeier rule, which states that solar cycles with shorter rise times (i.e., smaller $Td_{k}$) tend to have larger amplitudes (corresponding to smaller $Ts_{k}$). Lastly, $T0_k$ is the start time and $\tau_k$ is the duration of cycle $k$. 
Using the data composite (1874 -- 2023 E.C.) from \citet{mandal2020}, we obtain the values of the parameters $Td_{k}$ and $Ts_{k}$ for the thirteen solar cycles spanning Cycles 12 to 24.\\

\item \textit{Determination of a Correlation Between Cycle Parameters and the Open Solar Flux} \\

We use the same open flux composite $F_{o}$ used in Penza et al. 2024, constructed through inter-calibration of the following datasets: Open Solar Flux $F_{o}$ by \cite{usoskin2021} for the period 971 – 1899 E.C., Solar Modulation Potential by \cite{muscheler2007} for the period 1513-2001 E.C. and Solar Modulation Potential from a Neutron Monitor by \cite{usoskin2017} for the period 1950- 2020 E.C..\\
The relation identified in \cite{penza2024} establishes the following connection between the 11-year cycle-averaged $F_o$ values ($\overline{F_{o}}$) and the parameter  $P_k \equiv (Td_k/Ts_k)^2$, which is proportional to the analytical integral of Eq. \ref{cycle_form}:
  \begin{equation}
  \label{Pk_Fopen}
  P_{k} = (0.002  \pm 0.001 ) {\overline{F_{o}}} +  (-0.004 \pm  0.003)    
  \end{equation}

Eq. \ref{Pk_Fopen} allows us to obtain the values of $P_{k}$, and then of $Td_{k}$ and $Ts_{k}$ and to reconstruct the sunspot area from 971 E.C. 
This preliminary reconstruction of sunspot areas is necessary in order to reconstruct the plage coverage. It is needed because the most recent composite indicates a non-zero plage coverage during solar minima, making functional forms like ours unsuitable, as they are by definition zero at the beginning of each cycle.\\

\item \textit{Reconstruction of Plage Coverage} \\

The reconstruction of the plage coverage is derived from the well-established power-law relationship between plage and sunspot coverages \citep{chatzistergos22, penza2024}. In particular, in \cite{penza2024} the authors find the following relation between the monthly averages of sunspot \citep{mandal2020} and plage \citep{chatzistergos20} area coverages within the common time range 1903-2023 C.E.:

\begin{equation}
\label{plage_vs_spot}
\alpha_f = (1.3 \pm 0.1) \alpha_s^{(0.61 \pm 0.02)} +  (0.0042 \pm 0.0005)   
\end{equation}

The final reconstruction of plage coverage from 971 E.C. to the present is presented in Fig. \ref{plage}.

\end{enumerate}

\subsection{UV Band Reconstruction} 
  \label{UV_rec}

In order to reconstruct the UV bands over the last millenium, we use Eq. \ref{delta_Flux_UV2}, with the contrast parameters obtained by the fits along the single cycles. Nevertheless, it is necessary to account for the long-term variations affecting the open magnetic field, that we associate with the quiet network component. Accordingly, the equation can be rewritten as follows:

\begin{equation}
\label{delta_rec_norm}
\delta F(t) = C_{n} \left[ 1 + F_{LT}(t) \right] N_{norm} + \alpha_{f}(t) \delta_{fn}
\end{equation}

where $F_{LT}(t)$ is a normalized long-term modulation function and $N_{norm}$ is a normalization parameter. $F_{LT}(t)$ is the same function computed in \cite{penza2024}. It is derived by decomposing the ${F_o}$ composite into Intrinsic Mode Functions (IMFs) using Empirical Mode Decomposition (EMD) \citep{Huang1998}, and then selecting only those components with characteristic timescales longer than 22 years, along with the monotonic trend. In other words, 
this process removes the variability on timescales shorter than 22 years from ${F_o}$, which is already captured in the temporal evolution of the plage component. \\

\begin{figure}[h]
\centering
\includegraphics[width=\textwidth]{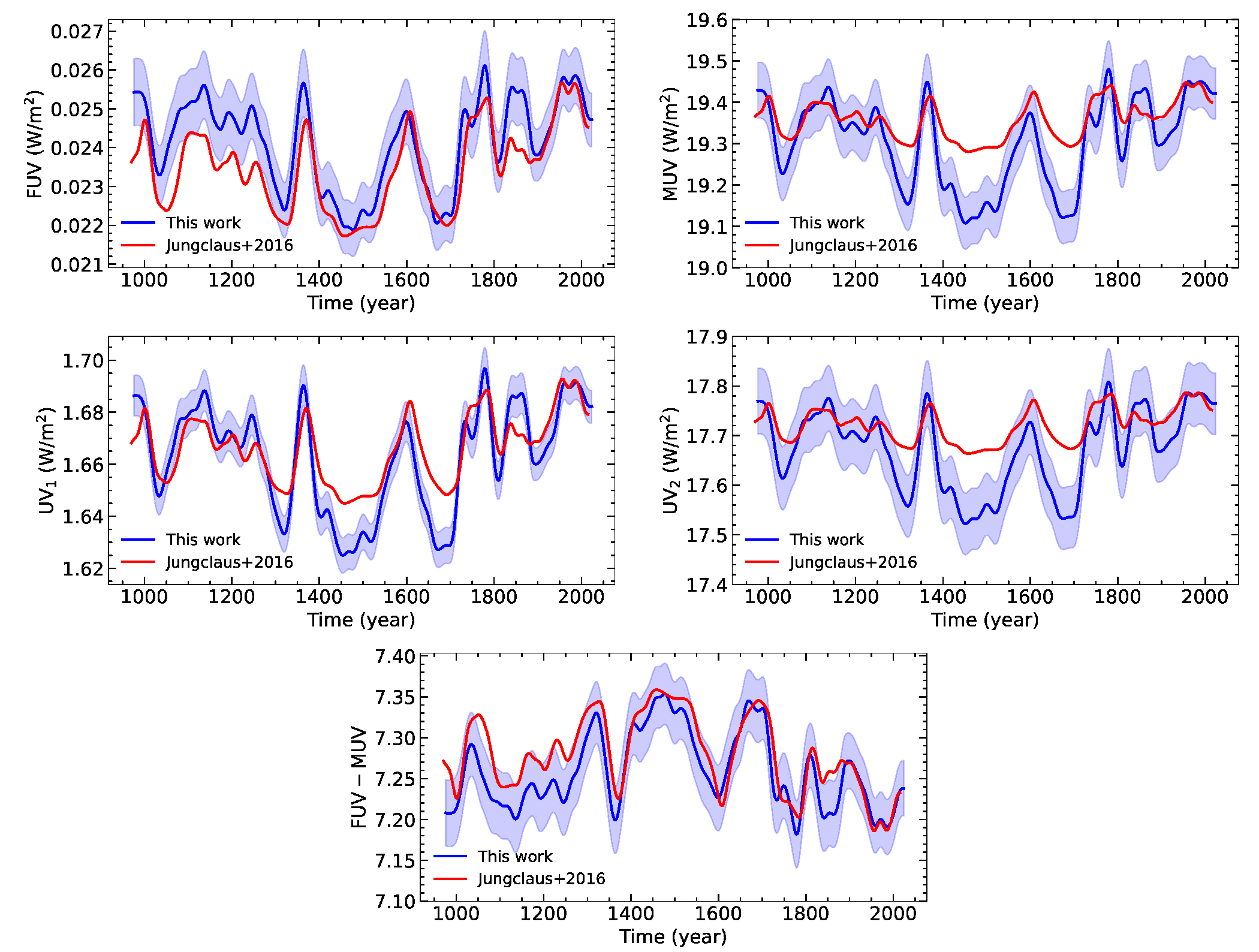}
\caption{Smoothed 22-year reconstructions from 971 CE to the present, obtained in this work (blue line), along with their uncertainty ranges (shaded blue regions). This uncertainty arises from the propagation of errors in the plage coverage and in the coefficients in Tab.\ref{tab2}.
The red lines represent the corresponding reconstructions from \cite{jungclaus2016}. The five panels show: FUV (top-left), MUV (top-right), $\mathrm{UV_{1}}$ (middle-left), $\mathrm{UV_{2}}$ (middle-right) and the color index FUV-MUV (bottom).
}
\label{rec_1000}
\end{figure}

For the final reconstruction, the two free parameters $N_{norm}$ and $F_{q}$ are obtained by the four best fit over the SSI3 composite (one for each wavelength band), by considering the whole period 1980-2020. The values of the free parameters are reported in Table \ref{tab2}, while the resulting final reconstructions of the four UV bands are shown in Fig. \ref{rec_1000}, which also includes the reconstruction of the FUV–MUV color index.

\begin{table}
\caption{Values of the parameters of Eq.\ref{delta_rec_norm}, obtained by best fitting to the full SSI3 dataset over the period 1980–2015.}
\label{tab2}
\begin{tabular}{lll}     
\hline                     
Band   &   $N_{norm}$ & $F_{q}$  ($W m^{-2}$) \\ 
\hline

FUV & 6.34 $\pm$ 0.01  &  0.020 $\pm$ 0.001 \\
MUV &   0.97 $\pm$ 0.01  &  31.10 $\pm$  0.01 \\
$UV_{1}$ &   2.16 $\pm$ 0.01 &  1.63 $\pm$ 0.01 \\
$UV_{2}$ &  0.91 $\pm$ 0.01  & 17.50 $\pm$ 0.01 \\
\hline
\end{tabular}
\end{table}

\section{Discussion and Conclusions} 
\label{S-Conclusion} 

In this work, we present a reconstruction of solar UV variability over the last millennium, covering the period from 971 to 2020 CE. The methodology used is the same as in \cite{penza2024}, and the reconstruction is based on plage coverage data from \cite{chatzistergos20} and the open magnetic field proxy from \cite{usoskin2021}. Using this approach, we reconstruct four UV bands: FUV (115–180~nm), MUV (180–310~nm), $\mathrm{UV_{1}}$ (100–243~nm), and $\mathrm{UV_{2}}$ (243–308~nm). In addition to these four bands, we also reconstruct the FUV–MUV color index.

\begin{figure}[h]
\centering
\includegraphics[width=\textwidth]{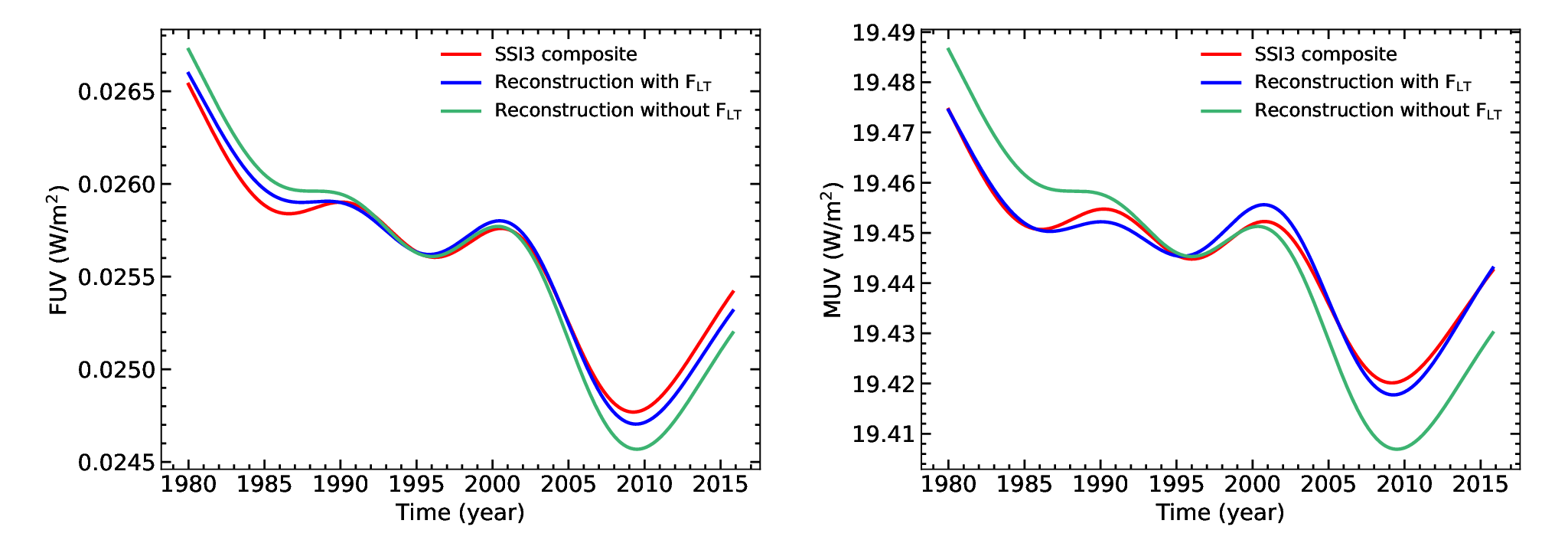}
\caption{Comparison between the reconstruction with (blue) and without (green) the long-term modulation $\mathrm{F_{LT}}$ and the SSI3 composite (red), for the FUV (left panel) and MUV (right panel) bands. All data are smoothed over 11 years.}
\label{MOD_noMOD}
\end{figure}
We have confirmed that including the long-term modulation factor $F_{LT}$ in the reconstruction improves its agreement with the SSI3 composite data over the 1980–2015 period, as clearly shown in Fig.~\ref{MOD_noMOD}. The figure presents results for the FUV and MUV bands, while similar improvements are also observed for the $\mathrm{UV_{1}}$ and $\mathrm{UV_{2}}$ bands (not shown). In particular, the long-term modulation factor plays a more significant role for the longer-wavelength bands (i.e., MUV) than for the shorter ones.
\begin{figure}[h]
\centering
\includegraphics[width=\textwidth]{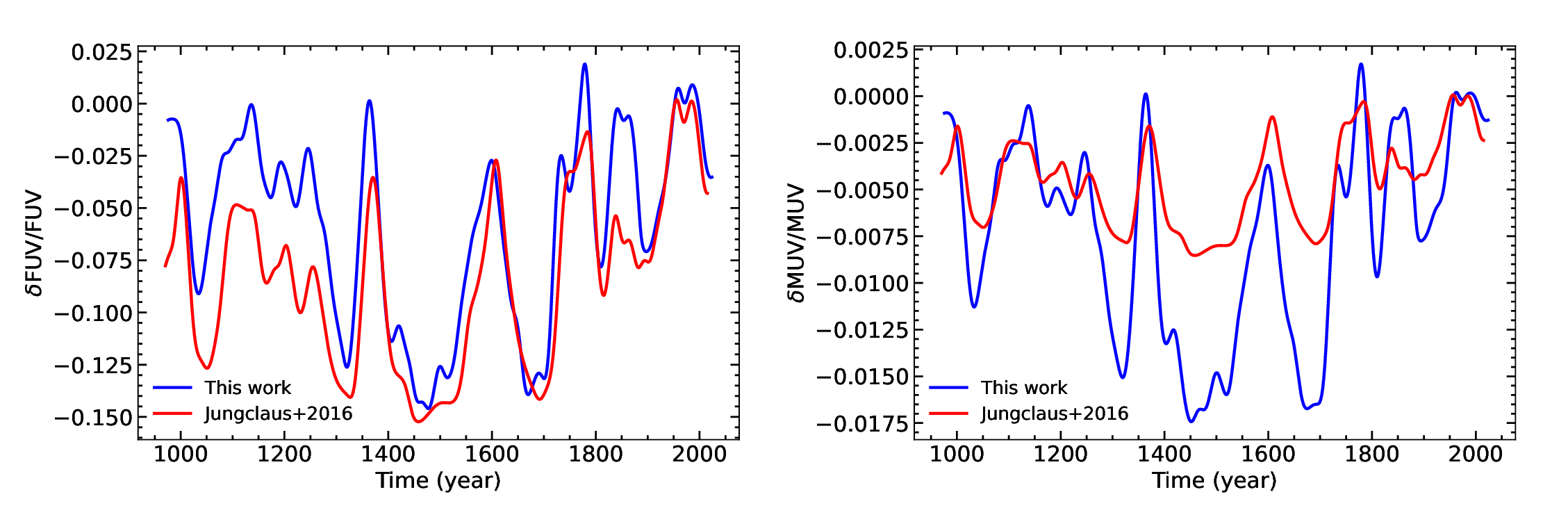}
\caption{Comparison of irradiance variations relative to the present in the FUV (left) and MUV (right) bands between our reconstruction and the SATIRE model \citep{jungclaus2016}. The present-day value is defined as the average of the SSI3 data from 1980 to 2015.}
\label{deltaUV_1000}
\end{figure}
The reconstruction includes two bands, here indicated with $UV_{1}$ and $UV_{2}$ which are directly associated with the photochemical production and destruction of ozone in the Earth's atmosphere.\\
The final reconstructed time series for each band are presented in Fig.~\ref{rec_1000}.
For the FUV and $\mathrm{UV_{1}}$ bands, our reconstruction closely resembles that of \cite{jungclaus2016}, which is based on the $^{14}$C record and SATIRE models.
Conversely, significant deviations are observed in the longer-wavelength bands (MUV and UV2).\\
Both our reconstruction and that of \cite{jungclaus2016} estimate a variation of approximately $15\%$ in the FUV band during the Spörer and Maunder minima relative to the present-day level, defined as the average value of the SSI3 data from 1980 to 2015.
However, the variation in the MUV band is considerably smaller, with the SATIRE model estimating only about $0.7\%$, whereas our reconstruction exceeds $1.6\%$ (Fig.~\ref{deltaUV_1000}).
This discrepancy likely arises from the modulation factor we introduced to the quiet-network component, which, as shown in Fig.~\ref{MOD_noMOD}, has a stronger influence on the longer wavelength bands than on the shorter ones.\\
The reconstructed FUV and MUV bands also allow for the calculation of the FUV–MUV color index, which can be used to characterize stellar UV emission \citep{Schindelm15, Calabro2021} and is strongly correlated with the Mg~II index  \citep{lovric2017, Criscuoli2018}. Because the FUV-MUV modulation is mainly driven by variations in the FUV, our reconstruction does not deviate significantly from the SATIRE model one. The computation of the FUV-MUV color index places this study within a broader stellar context. Studying UV irradiance variations in stars other than the Sun is a topic of great interest in stellar astrophysics \citep{Lubin2018, Namekata2023}. In fact, ultraviolet radiation plays a key role in shaping exoplanetary atmospheres through photochemical processes and atmospheric erosion, thereby influencing climate and surface conditions critical to habitability \citep[e.g.,][]{Segura2005,Reda2022, Spinelli2023, Li2024}. By characterizing solar UV variability over long timescales in relation to broad levels of solar activity, we gain valuable insights that can be extended to the study of stellar environments and their impact on planetary systems conditions.



\begin{acks}
The National Solar Observatory is operated by the Association of Universities for Research in Astronomy, Inc. (AURA), under cooperative agreement with the National Science Foundation. L.B. and S.C. are members of the international team on Reconstructing Solar and Heliospheric Magnetic Field Evolution Over the Past Century supported by the International Space Science Institute (ISSI), Bern, Switzerland.
 L.L. is supported by the Joint Research PhD Program in “Astronomy, Astrophysics and Space Science” between the University of Rome "Tor Vergata", the University of Rome "La Sapienza" and INAF. The authors acknowledge the Space It Up project funded by the Italian Space Agency, ASI, and the Ministry of University and Research, MUR, under contract n. 2024-5-E.0 - CUP n. I53D24000060005. “Realizzazione di un’Infrastruttura HW e SW (IHS)
presso il CGS/Matera (CUP F83C22002460005), in attuazione
del Piano Operativo del sub-investimento M1C2.I4.4 'In-Orbit
Economy - SST—FlyEye' del PNRR-FC”
\end{acks}







\begin{ethics}
\begin{conflict}
The authors declare no competing interests.
\end{conflict}
\end{ethics}

\bibliographystyle{spr-mp-sola}
\bibliography{sola_bibliography_example}  

\begin{thebibliography}{67}
\ifx\bisbn     \undefined \def\bisbn  #1{ISBN #1}\fi
\ifx\binits    \undefined \def\binits#1{#1}\fi
\ifx\bauthor   \undefined \def\bauthor#1{#1}\fi
\ifx\batitle   \undefined \def\batitle#1{#1}\fi
\ifx\bjtitle   \undefined \def\bjtitle#1{\textit{#1}}\fi
\ifx\bvolume   \undefined \def\bvolume#1{\textbf{#1}}\fi
\ifx\byear     \undefined \def\byear#1{#1}\fi
\ifx\bissue    \undefined \def\bissue#1{#1}\fi
\ifx\bfpage    \undefined \def\bfpage#1{#1}\fi
\ifx\blpage    \undefined \def\blpage #1{#1}\fi
\ifx\burl      \undefined \def\burl#1{#1}\fi
\ifx\href      \undefined \def\href#1#2{#2}\fi
\ifx\betal     \undefined \def\betal{et al.}\fi
\ifx\bctitle   \undefined \def\bctitle#1{#1}\fi
\ifx\beditor   \undefined \def\beditor#1{#1}\fi
\ifx\bbtitle   \undefined \def\bbtitle#1{\textit{#1}}\fi
\ifx\bedition  \undefined \def\bedition#1{#1}\fi
\ifx\bseriesno \undefined \def\bseriesno#1{\textbf{#1}}\fi
\ifx\blocation \undefined \def\blocation#1{#1}\fi
\ifx\bsertitle \undefined \def\bsertitle#1{\textit{#1}}\fi
\ifx\bsnm      \undefined \def\bsnm#1{#1}\fi
\ifx\bsuffix   \undefined \def\bsuffix#1{#1}\fi
\ifx\bparticle \undefined \def\bparticle#1{#1}\fi
\ifx\barticle  \undefined \def\barticle#1{}\fi
\ifx\binstitute  \undefined \def\binstitute#1{#1}\fi
\ifx\bpublisher  \undefined \def\bpublisher#1{#1}\fi
\ifx\doiurl    \undefined \def\doiurl#1{\href{#1}{DOI}}\fi
\makeatletter
\def\safeHref#1#2#3{\in@{http}{#2}\ifin@\href{#2}{#3}\else\href{#1#2}{#3}\fi}
\makeatother
\ifx\adsurl    \undefined
  \def\adsurl#1{\safeHref{https://ui.adsabs.harvard.edu/abs/}{#1}{ADS}}\fi
\ifx\arxivurl  \undefined
  \def\arxivurl#1{\safeHref{http://arxiv.org/abs/}{#1}{arXiv}}\fi
\ifx\botherref \undefined \def\botherref#1{}\fi
\ifx\url       \undefined \def\url#1{#1}\fi
\ifx\bchapter  \undefined \def\bchapter#1{}\fi
\ifx\bbook     \undefined \def\bbook#1{}\fi
\ifx\bcomment  \undefined \def\bcomment#1{#1}\fi
\ifx\oauthor   \undefined \def\oauthor#1{#1}\fi
\ifx\citeauthoryear \undefined\def \citeauthoryear#1{#1}\fi
\def\endbibitem {}
\ifx\bconflocation  \undefined \def\bconflocation#1{#1} \fi

\bibitem[\protect\citeauthoryear{{Ball} et~al.}{2011}]{ball2011}
\begin{barticle}
\bauthor{\bsnm{{Ball}}, \binits{W.T.}},
\bauthor{\bsnm{{Unruh}}, \binits{Y.C.}},
\bauthor{\bsnm{{Krivova}}, \binits{N.A.}},
\bauthor{\bsnm{{Solanki}}, \binits{S.}},
\bauthor{\bsnm{{Harder}}, \binits{J.W.}}:
\byear{2011},
\batitle{{Solar irradiance variability: a six-year comparison between SORCE
  observations and the SATIRE model}}.
\bjtitle{\aap}
\bvolume{530},
\bfpage{A71}.
\doiurl{https://doi.org/10.1051/0004-6361/201016189}.
\adsurl{2011A&A...530A..71B}.
\end{barticle}
\endbibitem

\bibitem[\protect\citeauthoryear{{Berrilli} et~al.}{1999}]{Berrilli1999}
\begin{barticle}
\bauthor{\bsnm{{Berrilli}}, \binits{F.}},
\bauthor{\bsnm{{Ermolli}}, \binits{I.}},
\bauthor{\bsnm{{Florio}}, \binits{A.}},
\bauthor{\bsnm{{Pietropaolo}}, \binits{E.}}:
\byear{1999},
\batitle{{Average properties and temporal variations of the geometry of solar
  network cells}}.
\bjtitle{\aap}
\bvolume{344},
\bfpage{965}.
\adsurl{1999A\%26A...344..965B}.
\end{barticle}
\endbibitem

\bibitem[\protect\citeauthoryear{{Berrilli} et~al.}{2020}]{berrilli2020}
\begin{barticle}
\bauthor{\bsnm{{Berrilli}}, \binits{F.}},
\bauthor{\bsnm{{Criscuoli}}, \binits{S.}},
\bauthor{\bsnm{{Penza}}, \binits{V.}},
\bauthor{\bsnm{{Lovric}}, \binits{M.}}:
\byear{2020},
\batitle{{Long-term (1749-2015) Variations of Solar UV Spectral Indices}}.
\bjtitle{\solphys}
\bvolume{295},
\bfpage{38}.
\doiurl{https://doi.org/10.1007/s11207-020-01603-5}.
\adsurl{2020SoPh..295...38B}.
\end{barticle}
\endbibitem

\bibitem[\protect\citeauthoryear{{Bigazzi}, {Cauli}, and
  {Berrilli}}{2020}]{Bigazzi2020}
\begin{barticle}
\bauthor{\bsnm{{Bigazzi}}, \binits{A.}},
\bauthor{\bsnm{{Cauli}}, \binits{C.}},
\bauthor{\bsnm{{Berrilli}}, \binits{F.}}:
\byear{2020},
\batitle{{Lower-thermosphere response to solar activity: an
  empirical-mode-decomposition analysis of GOCE 2009-2012 data}}.
\bjtitle{Annales Geophysicae}
\bvolume{38},
\bfpage{789}.
\doiurl{https://doi.org/10.5194/angeo-38-789-2020}.
\adsurl{2020AnGeo..38..789B}.
\end{barticle}
\endbibitem

\bibitem[\protect\citeauthoryear{{Bordi}, {Berrilli}, and
  {Pietropaolo}}{2015}]{Bordi2015}
\begin{barticle}
\bauthor{\bsnm{{Bordi}}, \binits{I.}},
\bauthor{\bsnm{{Berrilli}}, \binits{F.}},
\bauthor{\bsnm{{Pietropaolo}}, \binits{E.}}:
\byear{2015},
\batitle{{Long-term response of stratospheric ozone and temperature to solar
  variability}}.
\bjtitle{Annales Geophysicae}
\bvolume{33},
\bfpage{267}.
\doiurl{https://doi.org/10.5194/angeo-33-267-2015}.
\adsurl{2015AnGeo..33..267B}.
\end{barticle}
\endbibitem

\bibitem[\protect\citeauthoryear{{Calabr{\`o}} et~al.}{2021}]{Calabro2021}
\begin{barticle}
\bauthor{\bsnm{{Calabr{\`o}}}, \binits{A.}},
\bauthor{\bsnm{{Castellano}}, \binits{M.}},
\bauthor{\bsnm{{Pentericci}}, \binits{L.}},
\bauthor{\bsnm{{Fontanot}}, \binits{F.}},
\bauthor{\bsnm{{Menci}}, \binits{N.}},
\bauthor{\bsnm{{Cullen}}, \binits{F.}},
\bauthor{\bsnm{{McLure}}, \binits{R.}},
\bauthor{\bsnm{{Bolzonella}}, \binits{M.}},
\bauthor{\bsnm{{Cimatti}}, \binits{A.}},
\bauthor{\bsnm{{Marchi}}, \binits{F.}},
\bauthor{\bsnm{{Talia}}, \binits{M.}},
\bauthor{\bsnm{{Amor{\'\i}n}}, \binits{R.}},
\bauthor{\bsnm{{Cresci}}, \binits{G.}},
\bauthor{\bsnm{{De Lucia}}, \binits{G.}},
\bauthor{\bsnm{{Fynbo}}, \binits{J.}},
\bauthor{\bsnm{{Fontana}}, \binits{A.}},
\bauthor{\bsnm{{Franco}}, \binits{M.}},
\bauthor{\bsnm{{Hathi}}, \binits{N.P.}},
\bauthor{\bsnm{{Hibon}}, \binits{P.}},
\bauthor{\bsnm{{Hirschmann}}, \binits{M.}},
\bauthor{\bsnm{{Mannucci}}, \binits{F.}},
\bauthor{\bsnm{{Santini}}, \binits{P.}},
\bauthor{\bsnm{{Saxena}}, \binits{A.}},
\bauthor{\bsnm{{Schaerer}}, \binits{D.}},
\bauthor{\bsnm{{Xie}}, \binits{L.}},
\bauthor{\bsnm{{Zamorani}}, \binits{G.}}:
\byear{2021},
\batitle{{The VANDELS survey: The relation between the UV continuum slope and
  stellar metallicity in star-forming galaxies at z {\ensuremath{\sim}} 3}}.
\bjtitle{\aap}
\bvolume{646},
\bfpage{A39}.
\doiurl{https://doi.org/10.1051/0004-6361/202039244}.
\adsurl{2021A&A...646A..39C}.
\end{barticle}
\endbibitem

\bibitem[\protect\citeauthoryear{{Cebula}, {Park}, and
  {Heath}}{1988}]{Cebula1988}
\begin{barticle}
\bauthor{\bsnm{{Cebula}}, \binits{R.P.}},
\bauthor{\bsnm{{Park}}, \binits{H.}},
\bauthor{\bsnm{{Heath}}, \binits{D.F.}}:
\byear{1988},
\batitle{{Characterization of the Nimbus-7 SBUV radiometer for the long-term
  monitoring of stratospheric ozone}}.
\bjtitle{Journal of Atmospheric and Oceanic Technology}
\bvolume{5},
\bfpage{215}.
\doiurl{https://doi.org/10.1175/1520-0426(1988)005<0215:COTNSR>2.0.CO;2}.
\adsurl{1988JAtOT...5..215C}.
\end{barticle}
\endbibitem

\bibitem[\protect\citeauthoryear{{Chapman}}{1930}]{chapman1930}
\begin{barticle}
\bauthor{\bsnm{{Chapman}}, \binits{S.}}:
\byear{1930},
\batitle{{A theory of upper atmosphere ozone}}.
\bjtitle{Mem. R. Metrol. Soc.}
\bvolume{3},
\bfpage{103}.
\end{barticle}
\endbibitem

\bibitem[\protect\citeauthoryear{{Chatzistergos}
  et~al.}{2020}]{chatzistergos20}
\begin{barticle}
\bauthor{\bsnm{{Chatzistergos}}, \binits{T.}},
\bauthor{\bsnm{{Ermolli}}, \binits{I.}},
\bauthor{\bsnm{{Krivova}}, \binits{N.A.}},
\bauthor{\bsnm{K.}, \binits{S.S.}},
\bauthor{\bsnm{{Banerje}}, \binits{D.}},
\bauthor{\bsnm{{Barata}}, \binits{T.}},
\bauthor{\bsnm{{Belik}}, \binits{R.} \bsuffix{M.m~{Gaifera}}},
\bauthor{\bsnm{{Garcia, A.}}},
\bauthor{\bsnm{{Hanaoka}}, \binits{Y.}},
\bauthor{\bsnm{{Hedge}}, \binits{M.}},
\bauthor{\bsnm{{Klimes}}, \binits{J.}},
\bauthor{\bsnm{{Korokhin}}, \binits{V.V.}},
\bauthor{\bsnm{{Lourenço}}, \binits{A.}},
\bauthor{\bsnm{{Malherbe11}}, \binits{J.}},
\bauthor{\bsnm{{Marchenko}}, \binits{G.}},
\bauthor{\bsnm{{Peixinho}}, \binits{N.}},
\bauthor{\bsnm{{Sakurai}}, \binits{T.}},
\bauthor{\bsnm{{Tlatov}}, \binits{A.}}:
\byear{2020},
\batitle{{Analysis of full-disc Ca II K spectroheliograms. III. Plage area
  composite series covering 1892-2019}}.
\bjtitle{\aap}
\bvolume{639},
\bfpage{22}.
\doiurl{https://doi.org/10.1051/0004-6361/202037746}.
\end{barticle}
\endbibitem

\bibitem[\protect\citeauthoryear{{Chatzistergos}
  et~al.}{2022}]{chatzistergos22}
\begin{barticle}
\bauthor{\bsnm{{Chatzistergos}}, \binits{T.}},
\bauthor{\bsnm{{Ermolli}}, \binits{I.}},
\bauthor{\bsnm{{Krivova}}, \binits{N.A.}},
\bauthor{\bsnm{{Barata}}, \binits{T.}},
\bauthor{\bsnm{{Carvalho}}, \binits{S.}},
\bauthor{\bsnm{{Malherbe11}}, \binits{J.}}:
\byear{2022},
\batitle{{Scrutinising the relationship between plage areas and sunspot areas
  and numbers}}.
\bjtitle{\aap}
\bvolume{667},
\bfpage{21}.
\doiurl{https://doi.org/10.1051/0004-6361/202244913}.
\end{barticle}
\endbibitem

\bibitem[\protect\citeauthoryear{{Criscuoli}}{2016}]{Criscuoli2016}
\begin{barticle}
\bauthor{\bsnm{{Criscuoli}}, \binits{S.}}:
\byear{2016},
\batitle{{Angular Dependence of the Facular-Sunspot Coverage Relation as
  Derived by MDI Magnetograms}}.
\bjtitle{\solphys}
\bvolume{291},
\bfpage{1957}.
\doiurl{https://doi.org/10.1007/s11207-016-0947-5}.
\adsurl{2016SoPh..291.1957C}.
\end{barticle}
\endbibitem

\bibitem[\protect\citeauthoryear{{Criscuoli}}{2019}]{criscuoli2019}
\begin{barticle}
\bauthor{\bsnm{{Criscuoli}}, \binits{S.}}:
\byear{2019},
\batitle{{Effects of Continuum Fudging on Non-LTE Synthesis of Stellar Spectra.
  I. Effects on Estimates of UV Continua and Solar Spectral Irradiance
  Variability}}.
\bjtitle{\apj}
\bvolume{872},
\bfpage{52}.
\doiurl{https://doi.org/10.3847/1538-4357/aaf6b7}.
\adsurl{2019ApJ...872...52C}.
\end{barticle}
\endbibitem

\bibitem[\protect\citeauthoryear{{Criscuoli} et~al.}{2018}]{Criscuoli2018}
\begin{barticle}
\bauthor{\bsnm{{Criscuoli}}, \binits{S.}},
\bauthor{\bsnm{{Penza}}, \binits{V.}},
\bauthor{\bsnm{{Lovric}}, \binits{M.}},
\bauthor{\bsnm{{Berrilli}}, \binits{F.}}:
\byear{2018},
\batitle{{The Correlation of Synthetic UV Color versus Mg II Index along the
  Solar Cycle}}.
\bjtitle{\apj}
\bvolume{865},
\bfpage{22}.
\doiurl{https://doi.org/10.3847/1538-4357/aad809}.
\adsurl{2018ApJ...865...22C}.
\end{barticle}
\endbibitem

\bibitem[\protect\citeauthoryear{{Criscuoli} et~al.}{2023}]{Criscuoli2023}
\begin{barticle}
\bauthor{\bsnm{{Criscuoli}}, \binits{S.}},
\bauthor{\bsnm{{Marchenko}}, \binits{S.}},
\bauthor{\bsnm{{DeLand}}, \binits{M.}},
\bauthor{\bsnm{{Choudhary}}, \binits{D.}},
\bauthor{\bsnm{{Kopp}}, \binits{G.}}:
\byear{2023},
\batitle{{Understanding Sun-as-a-Star Variability of Solar Balmer Lines}}.
\bjtitle{\apj}
\bvolume{951},
\bfpage{151}.
\doiurl{https://doi.org/10.3847/1538-4357/acd17d}.
\adsurl{2023ApJ...951..151C}.
\end{barticle}
\endbibitem

\bibitem[\protect\citeauthoryear{{Devi} et~al.}{2021}]{pooja2021}
\begin{barticle}
\bauthor{\bsnm{{Devi}}, \binits{P.}},
\bauthor{\bsnm{{Singh}}, \binits{J.}},
\bauthor{\bsnm{{Chandra}}, \binits{R.}},
\bauthor{\bsnm{{Priyal}}, \binits{M.}},
\bauthor{\bsnm{{Joshi}}, \binits{R.}}:
\byear{2021},
\batitle{{Variation of Chromospheric Features as a Function of Latitude and
  Time Using Ca-K Spectroheliograms for Solar Cycles 15 - 23: Implications for
  Meridional Flow}}.
\bjtitle{\solphys}
\bvolume{296},
\bfpage{49}.
\doiurl{https://doi.org/10.1007/s11207-021-01798-1}.
\adsurl{2021SoPh..296...49D}.
\end{barticle}
\endbibitem

\bibitem[\protect\citeauthoryear{{Ermolli}, {Berrilli}, and
  {Florio}}{2003}]{Ermolli2003}
\begin{barticle}
\bauthor{\bsnm{{Ermolli}}, \binits{I.}},
\bauthor{\bsnm{{Berrilli}}, \binits{F.}},
\bauthor{\bsnm{{Florio}}, \binits{A.}}:
\byear{2003},
\batitle{{A measure of the network radiative properties over the solar activity
  cycle}}.
\bjtitle{\aap}
\bvolume{412},
\bfpage{857}.
\doiurl{https://doi.org/10.1051/0004-6361:20031479}.
\adsurl{2003A\%26A...412..857E}.
\end{barticle}
\endbibitem

\bibitem[\protect\citeauthoryear{{Ermolli}, {Giorgi}, and
  {Chatzistergos}}{2022}]{ermolli2022}
\begin{barticle}
\bauthor{\bsnm{{Ermolli}}, \binits{I.}},
\bauthor{\bsnm{{Giorgi}}, \binits{F.}},
\bauthor{\bsnm{{Chatzistergos}}, \binits{T.}}:
\byear{2022},
\batitle{{Rome Precision Solar Photometric Telescope: precision solar full-disk
  photometry during solar cycles 23{\textendash}25}}.
\bjtitle{Frontiers in Astronomy and Space Sciences}
\bvolume{9},
\bfpage{352}.
\doiurl{https://doi.org/10.3389/fspas.2022.1042740}.
\adsurl{2022FrASS...942740E}.
\end{barticle}
\endbibitem

\bibitem[\protect\citeauthoryear{{Fontenla} et~al.}{1999}]{fontenla1999}
\begin{barticle}
\bauthor{\bsnm{{Fontenla}}, \binits{J.}},
\bauthor{\bsnm{{White}}, \binits{O.R.}},
\bauthor{\bsnm{{Fox}}, \binits{P.A.}},
\bauthor{\bsnm{{Avrett}}, \binits{E.H.}},
\bauthor{\bsnm{{Kurucz}}, \binits{R.L.}}:
\byear{1999},
\batitle{{Calculation of Solar Irradiances. I. Synthesis of the Solar
  Spectrum}}.
\bjtitle{\apj}
\bvolume{518},
\bfpage{480}.
\doiurl{https://doi.org/10.1086/307258}.
\adsurl{1999ApJ...518..480F}.
\end{barticle}
\endbibitem

\bibitem[\protect\citeauthoryear{{Fontenla} et~al.}{2011}]{fontenla2011}
\begin{barticle}
\bauthor{\bsnm{{Fontenla}}, \binits{J.M.}},
\bauthor{\bsnm{{Harder}}, \binits{J.}},
\bauthor{\bsnm{{Livingston}}, \binits{W.}},
\bauthor{\bsnm{{Snow}}, \binits{M.}},
\bauthor{\bsnm{{Woods}}, \binits{T.}}:
\byear{2011},
\batitle{{High-resolution solar spectral irradiance from extreme ultraviolet to
  far infrared}}.
\bjtitle{Journal of Geophysical Research (Atmospheres)}
\bvolume{116},
\bfpage{D20108}.
\doiurl{https://doi.org/10.1029/2011JD016032}.
\adsurl{2011JGRD..11620108F}.
\end{barticle}
\endbibitem

\bibitem[\protect\citeauthoryear{{Fr{\"o}hlich} and
  {Lean}}{2004}]{frohlich2004}
\begin{barticle}
\bauthor{\bsnm{{Fr{\"o}hlich}}, \binits{C.}},
\bauthor{\bsnm{{Lean}}, \binits{J.}}:
\byear{2004},
\batitle{{Solar radiative output and its variability: evidence and
  mechanisms}}.
\bjtitle{\aapr}
\bvolume{12},
\bfpage{273}.
\doiurl{https://doi.org/10.1007/s00159-004-0024-1}.
\adsurl{2004A&ARv..12..273F}.
\end{barticle}
\endbibitem

\bibitem[\protect\citeauthoryear{{Galuzzo} et~al.}{2021}]{Galuzzo2021}
\begin{barticle}
\bauthor{\bsnm{{Galuzzo}}, \binits{D.}},
\bauthor{\bsnm{{Cagnazzo}}, \binits{C.}},
\bauthor{\bsnm{{Berrilli}}, \binits{F.}},
\bauthor{\bsnm{{Fierli}}, \binits{F.}},
\bauthor{\bsnm{{Giovannelli}}, \binits{L.}}:
\byear{2021},
\batitle{{Three-dimensional Climate Simulations for the Detectability of
  Proxima Centauri b}}.
\bjtitle{\apj}
\bvolume{909},
\bfpage{191}.
\doiurl{https://doi.org/10.3847/1538-4357/abdeb4}.
\adsurl{2021ApJ...909..191G}.
\end{barticle}
\endbibitem

\bibitem[\protect\citeauthoryear{{Haigh}}{2007}]{Haigh2007}
\begin{barticle}
\bauthor{\bsnm{{Haigh}}, \binits{J.D.}}:
\byear{2007},
\batitle{{The Sun and the Earth's Climate}}.
\bjtitle{Living Reviews in Solar Physics}
\bvolume{4},
\bfpage{2}.
\doiurl{https://doi.org/10.12942/lrsp-2007-2}.
\adsurl{2007LRSP....4....2H}.
\end{barticle}
\endbibitem

\bibitem[\protect\citeauthoryear{{Huang} et~al.}{1998}]{Huang1998}
\begin{barticle}
\bauthor{\bsnm{{Huang}}, \binits{N.E.}},
\bauthor{\bsnm{{Shen}}, \binits{Z.}},
\bauthor{\bsnm{{Long}}, \binits{S.R.}},
\bauthor{\bsnm{{Wu}}, \binits{M.C.}},
\bauthor{\bsnm{{Shih}}, \binits{H.H.}},
\bauthor{\bsnm{{Zheng}}, \binits{Q.}},
\bauthor{\bsnm{{Yen}}, \binits{N.-C.}},
\bauthor{\bsnm{{Tung}}, \binits{C.C.}},
\bauthor{\bsnm{{Liu}}, \binits{H.H.}}:
\byear{1998},
\batitle{{The empirical mode decomposition and the Hilbert spectrum for
  nonlinear and non-stationary time series analysis}}.
\bjtitle{Proceedings of the Royal Society of London Series A}
\bvolume{454},
\bfpage{903}.
\doiurl{https://doi.org/10.1098/rspa.1998.0193}.
\adsurl{1998RSPSA.454..903H}.
\end{barticle}
\endbibitem

\bibitem[\protect\citeauthoryear{{Jungclaus} et~al.}{2016}]{jungclaus2016}
\begin{barticle}
\bauthor{\bsnm{{Jungclaus}}, \binits{J.H.}},
\bauthor{\bsnm{{Bard}}, \binits{E.}},
\bauthor{\bsnm{{Baroni}}, \binits{M.}},
\bauthor{\bsnm{{Braconnot}}, \binits{P.}},
\bauthor{\bsnm{{Cao}}, \binits{J.}},
\bauthor{\bsnm{{Chini}}, \binits{L.P.}},
\bauthor{\bsnm{{Egorova}}, \binits{T.}},
\bauthor{\bsnm{{Evans}}, \binits{M.}},
\bauthor{\bsnm{{Fidel Gonz{\'a}lez-Rouco}}, \binits{J.}},
\bauthor{\bsnm{{Goosse}}, \binits{H.}},
\bauthor{\bsnm{{Hurtt}}, \binits{G.C.}},
\bauthor{\bsnm{{Joos}}, \binits{F.}},
\bauthor{\bsnm{{Kaplan}}, \binits{J.O.}},
\bauthor{\bsnm{{Khodri}}, \binits{M.}},
\bauthor{\bsnm{{Klein Goldewijk}}, \binits{K.}},
\bauthor{\bsnm{{Krivova}}, \binits{N.}},
\bauthor{\bsnm{{LeGrande}}, \binits{A.N.}},
\bauthor{\bsnm{{Lorenz}}, \binits{S.J.}},
\bauthor{\bsnm{{Luterbacher}}, \binits{J.}},
\bauthor{\bsnm{{Man}}, \binits{W.}},
\bauthor{\bsnm{{Maycock}}, \binits{A.C.}},
\bauthor{\bsnm{{Meinshausen}}, \binits{M.}},
\bauthor{\bsnm{{Moberg}}, \binits{A.}},
\bauthor{\bsnm{{Muscheler}}, \binits{R.}},
\bauthor{\bsnm{{Nehrbass-Ahles}}, \binits{C.}},
\bauthor{\bsnm{{Otto-Bliesner}}, \binits{B.I.}},
\bauthor{\bsnm{{Phipps}}, \binits{S.J.}},
\bauthor{\bsnm{{Pongratz}}, \binits{J.}},
\bauthor{\bsnm{{Rozanov}}, \binits{E.}},
\bauthor{\bsnm{{Schmidt}}, \binits{G.A.}},
\bauthor{\bsnm{{Schmidt}}, \binits{H.}},
\bauthor{\bsnm{{Schmutz}}, \binits{W.}},
\bauthor{\bsnm{{Schurer}}, \binits{A.}},
\bauthor{\bsnm{{Shapiro}}, \binits{A.I.}},
\bauthor{\bsnm{{Sigl}}, \binits{M.}},
\bauthor{\bsnm{{Smerdon}}, \binits{J.E.}},
\bauthor{\bsnm{{Solanki}}, \binits{S.K.}},
\bauthor{\bsnm{{Timmreck}}, \binits{C.}},
\bauthor{\bsnm{{Toohey}}, \binits{M.}},
\bauthor{\bsnm{{Usoskin}}, \binits{I.G.}},
\bauthor{\bsnm{{Wagner}}, \binits{S.}},
\bauthor{\bsnm{{Wu}}, \binits{C.-J.}},
\bauthor{\bsnm{{Leng Yeo}}, \binits{K.}},
\bauthor{\bsnm{{Zanchettin}}, \binits{D.}},
\bauthor{\bsnm{{Zhang}}, \binits{Q.}},
\bauthor{\bsnm{{Zorita}}, \binits{E.}}:
\byear{2016},
\batitle{{The PMIP4 contribution to CMIP6 - Part 3: The last millennium,
  scientific objective, and experimental design for the PMIP4 past1000
  simulations}}.
\bjtitle{Geoscientific Model Development}
\bvolume{10},
\bfpage{4005}.
\doiurl{https://doi.org/10.5194/gmd-10-4005-2017}.
\adsurl{2017GMD....10.4005J}.
\end{barticle}
\endbibitem

\bibitem[\protect\citeauthoryear{{Kakuwa} and {Satoru}}{2022}]{kakuwa}
\begin{barticle}
\bauthor{\bsnm{{Kakuwa}}, \binits{J.}},
\bauthor{\bsnm{{Satoru}}, \binits{U.}}:
\byear{2022},
\batitle{{Investigation of the Long-term Variation of Solar Ca II K Intensity.
  II. Reconstruction of Solar UV Irradiance}}.
\bjtitle{\aj}
\bvolume{928},
\bfpage{97}.
\doiurl{https://doi.org/10.3847/1538-4357/ac5963}.
\adsurl{https://iopscience.iop.org/article/10.3847/1538-4357/ac5963/pdf}.
\end{barticle}
\endbibitem

\bibitem[\protect\citeauthoryear{{Kopp} et~al.}{2016}]{kopp2016}
\begin{barticle}
\bauthor{\bsnm{{Kopp}}, \binits{G.}},
\bauthor{\bsnm{{Krivova}}, \binits{N.}},
\bauthor{\bsnm{{Wu}}, \binits{C.J.}},
\bauthor{\bsnm{{Lean}}, \binits{J.}}:
\byear{2016},
\batitle{{The Impact of the Revised Sunspot Record on Solar Irradiance
  Reconstructions}}.
\bjtitle{\solphys}
\bvolume{291},
\bfpage{2951}.
\doiurl{https://doi.org/10.1007/s11207-016-0853-x}.
\adsurl{2016SoPh..291.2951K}.
\end{barticle}
\endbibitem

\bibitem[\protect\citeauthoryear{{Krivova} et~al.}{2003}]{krivova2003}
\begin{barticle}
\bauthor{\bsnm{{Krivova}}, \binits{N.A.}},
\bauthor{\bsnm{{Solanki}}, \binits{S.K.}},
\bauthor{\bsnm{{Fligge}}, \binits{M.}},
\bauthor{\bsnm{{Unruh}}, \binits{Y.C.}}:
\byear{2003},
\batitle{{Reconstruction of solar irradiance variations in cycle 23: Is solar
  surface magnetism the cause?}}
\bjtitle{\aap}
\bvolume{399},
\bfpage{L1}.
\doiurl{https://doi.org/10.1051/0004-6361:20030029}.
\adsurl{2003A&A...399L...1K}.
\end{barticle}
\endbibitem

\bibitem[\protect\citeauthoryear{{Lean}}{1997}]{Lean1997b}
\begin{barticle}
\bauthor{\bsnm{{Lean}}, \binits{J.}}:
\byear{1997},
\batitle{{The Sun's Variable Radiation and Its Relevance For Earth}}.
\bjtitle{Annual Review of Astronomy and Astrophysics}
\bvolume{35},
\bfpage{33}.
\doiurl{https://doi.org/10.1146/annurev.astro.35.1.33}.
\adsurl{1997ARA&A..35...33L}.
\end{barticle}
\endbibitem

\bibitem[\protect\citeauthoryear{{Lean} et~al.}{1997}]{Lean1997}
\begin{barticle}
\bauthor{\bsnm{{Lean}}, \binits{J.L.}},
\bauthor{\bsnm{{Rottman}}, \binits{G.J.}},
\bauthor{\bsnm{{Kyle}}, \binits{H.L.}},
\bauthor{\bsnm{{Woods}}, \binits{T.N.}},
\bauthor{\bsnm{{Hickey}}, \binits{J.R.}},
\bauthor{\bsnm{{Puga}}, \binits{L.C.}}:
\byear{1997},
\batitle{{Detection and parameterization of variations in solar mid- and
  near-ultraviolet radiation (200-400 nm)}}.
\bjtitle{\jgr}
\bvolume{102},
\bfpage{29939}.
\doiurl{https://doi.org/10.1029/97JD02092}.
\adsurl{1997JGR...10229939L}.
\end{barticle}
\endbibitem

\bibitem[\protect\citeauthoryear{{Lean} et~al.}{2020}]{lean2020}
\begin{barticle}
\bauthor{\bsnm{{Lean}}, \binits{J.L.}},
\bauthor{\bsnm{{Coddington}}, \binits{O.}},
\bauthor{\bsnm{{Marchenko}}, \binits{S.V.}},
\bauthor{\bsnm{{Machol}}, \binits{J.}},
\bauthor{\bsnm{{DeLand}}, \binits{M.T.}},
\bauthor{\bsnm{{Kopp}}, \binits{G.}}:
\byear{2020},
\batitle{{Solar Irradiance Variability: Modeling the Measurements}}.
\bjtitle{Earth and Space Science}
\bvolume{7},
\bfpage{00645}.
\doiurl{https://doi.org/10.1029/2019EA000645}.
\adsurl{2020E&SS....700645L}.
\end{barticle}
\endbibitem

\bibitem[\protect\citeauthoryear{{Li} et~al.}{2024}]{Li2024}
\begin{barticle}
\bauthor{\bsnm{{Li}}, \binits{X.}},
\bauthor{\bsnm{{Wang}}, \binits{S.}},
\bauthor{\bsnm{{Han}}, \binits{H.}},
\bauthor{\bsnm{{Yang}}, \binits{H.}},
\bauthor{\bsnm{{Zheng}}, \binits{C.}},
\bauthor{\bsnm{{Huang}}, \binits{Y.}},
\bauthor{\bsnm{{Liu}}, \binits{J.}}:
\byear{2024},
\batitle{{Ultraviolet and Chromospheric Activity and Habitability of M Stars}}.
\bjtitle{\apj}
\bvolume{966},
\bfpage{69}.
\doiurl{https://doi.org/10.3847/1538-4357/ad3038}.
\adsurl{2024ApJ...966...69L}.
\end{barticle}
\endbibitem

\bibitem[\protect\citeauthoryear{{Lilensten} and
  {Tourpali}}{2015}]{lilensten2015}
\begin{bchapter}
\bauthor{\bsnm{{Lilensten}}, \binits{J.}},
\bauthor{\bsnm{{Tourpali}}, \binits{K.}}:
\byear{2015},
\bctitle{{Direct impact of solar irradiance variability}}.
In: \bbtitle{Earth’s climate response to a changing Sun},
\bpublisher{EDP Science},
\bfpage{229}.
\bcomment{Chap. 4.1}.
\bisbn{978--2--7598--1733--7}.
\doiurl{https://doi.org/10.1051/978-2-7598-1733-7.c1276}.
\end{bchapter}
\endbibitem

\bibitem[\protect\citeauthoryear{{Linsky}}{2014}]{linsky2014}
\begin{barticle}
\bauthor{\bsnm{{Linsky}}, \binits{J.}}:
\byear{2014},
\batitle{{The Radiation Environment of Exoplanet Atmospheres}}.
\bjtitle{Challenges}
\bvolume{5},
\bfpage{351}.
\doiurl{https://doi.org/10.3390/challe5020351}.
\adsurl{2014Chall...5..351L}.
\end{barticle}
\endbibitem

\bibitem[\protect\citeauthoryear{{Liu} et~al.}{2023}]{liu2023}
\begin{barticle}
\bauthor{\bsnm{{Liu}}, \binits{H.-L.}},
\bauthor{\bsnm{{Rempel}}, \binits{M.}},
\bauthor{\bsnm{{Danabasoglu}}, \binits{G.}},
\bauthor{\bsnm{{Solomon}}, \binits{S.C.}},
\bauthor{\bsnm{{McInerney}}, \binits{J.M.}}:
\byear{2023},
\batitle{{Climate Responses Under an Extreme Quiet Sun Scenario}}.
\bjtitle{Journal of Geophysical Research (Atmospheres)}
\bvolume{128},
\bfpage{e2022JD037626}.
\doiurl{https://doi.org/10.1029/2022JD037626}.
\adsurl{2023JGRD..12837626L}.
\end{barticle}
\endbibitem

\bibitem[\protect\citeauthoryear{{Lockwood}}{2012}]{lockwood2012}
\begin{barticle}
\bauthor{\bsnm{{Lockwood}}, \binits{M.}}:
\byear{2012},
\batitle{{Solar influence on global and regional climates}}.
\bjtitle{Surv. Geophys}
\bvolume{33},
\bfpage{503}.
\doiurl{https://doi.org/10.1007/s10712-012-9181-3}.
\end{barticle}
\endbibitem

\bibitem[\protect\citeauthoryear{{Lovric} et~al.}{2017}]{lovric2017}
\begin{barticle}
\bauthor{\bsnm{{Lovric}}, \binits{M.}},
\bauthor{\bsnm{{Tosone}}, \binits{F.}},
\bauthor{\bsnm{{Pietropaolo}}, \binits{E.}},
\bauthor{\bsnm{{Del Moro}}, \binits{D.}},
\bauthor{\bsnm{{Giovannelli}}, \binits{L.}},
\bauthor{\bsnm{{Cagnazzo}}, \binits{C.}},
\bauthor{\bsnm{{Berrilli}}, \binits{F.}}:
\byear{2017},
\batitle{{The dependence of the [FUV-MUV] colour on solar cycle}}.
\bjtitle{Journal of Space Weather and Space Climate}
\bvolume{7},
\bfpage{A6}.
\doiurl{https://doi.org/10.1051/swsc/2017001}.
\adsurl{2017JSWSC...7A...6L}.
\end{barticle}
\endbibitem

\bibitem[\protect\citeauthoryear{{Lubin}, {Melis}, and
  {Tytler}}{2018}]{Lubin2018}
\begin{barticle}
\bauthor{\bsnm{{Lubin}}, \binits{D.}},
\bauthor{\bsnm{{Melis}}, \binits{C.}},
\bauthor{\bsnm{{Tytler}}, \binits{D.}}:
\byear{2018},
\batitle{{Ultraviolet Flux Decrease Under a Grand Minimum from IUE
  Short-wavelength Observation of Solar Analogs}}.
\bjtitle{\apjl}
\bvolume{852},
\bfpage{L4}.
\doiurl{https://doi.org/10.3847/2041-8213/aaa124}.
\adsurl{2018ApJ...852L...4L}.
\end{barticle}
\endbibitem

\bibitem[\protect\citeauthoryear{{Mandal} et~al.}{2020}]{mandal2020}
\begin{barticle}
\bauthor{\bsnm{{Mandal}}, \binits{S.}},
\bauthor{\bsnm{{Krivova}}, \binits{N.A.}},
\bauthor{\bsnm{{Solanki}}, \binits{S.K.}},
\bauthor{\bsnm{{Sinha}}, \binits{N.}},
\bauthor{\bsnm{{Banerjee}}, \binits{D.}}:
\byear{2020},
\batitle{{Sunspot area catalog revisited: Daily cross-calibrated areas since
  1874}}.
\bjtitle{\aap}
\bvolume{640},
\bfpage{A78}.
\doiurl{https://doi.org/10.1051/0004-6361/202037547}.
\adsurl{2020A&A...640A..78M}.
\end{barticle}
\endbibitem

\bibitem[\protect\citeauthoryear{{Marchenko} et~al.}{2024}]{marchenko2024}
\begin{barticle}
\bauthor{\bsnm{{Marchenko}}, \binits{S.V.}},
\bauthor{\bsnm{{Ludewig}}, \binits{A.}},
\bauthor{\bsnm{{Criscuoli}}, \binits{S.}},
\bauthor{\bsnm{{Al Moulla}}, \binits{K.}},
\bauthor{\bsnm{{Choudhary}}, \binits{D.P.}},
\bauthor{\bsnm{{DeLand}}, \binits{M.T.}},
\bauthor{\bsnm{{Kopp}}, \binits{G.}},
\bauthor{\bsnm{{Loots}}, \binits{E.}},
\bauthor{\bsnm{{van der Plas}}, \binits{E.}},
\bauthor{\bsnm{{Veefkind}}, \binits{P.}}:
\byear{2024},
\batitle{{Sun-as-a-Star Spectral Line Variability in the 300{\textendash}2390
  nm Wavelength Range}}.
\bjtitle{\apj}
\bvolume{977},
\bfpage{33}.
\doiurl{https://doi.org/10.3847/1538-4357/ad888f}.
\adsurl{2024ApJ...977...33M}.
\end{barticle}
\endbibitem

\bibitem[\protect\citeauthoryear{{Matthes} et~al.}{2017}]{matthes2017}
\begin{barticle}
\bauthor{\bsnm{{Matthes}}, \binits{K.}},
\bauthor{\bsnm{{Funke}}, \binits{B.}},
\bauthor{\bsnm{{Andersson}}, \binits{M.E.}},
\bauthor{\bsnm{{Barnard}}, \binits{L.}},
\bauthor{\bsnm{{Beer}}, \binits{J.}},
\bauthor{\bsnm{{Charbonneau}}, \binits{P.}},
\bauthor{\bsnm{{Clilverd}}, \binits{M.A.}},
\bauthor{\bsnm{{Dudok de Wit}}, \binits{T.}},
\bauthor{\bsnm{{Haberreiter}}, \binits{M.}},
\bauthor{\bsnm{{Hendry}}, \binits{A.}},
\bauthor{\bsnm{{Jackman}}, \binits{C.H.}},
\bauthor{\bsnm{{Kretzschmar}}, \binits{M.}},
\bauthor{\bsnm{{Kruschke}}, \binits{T.}},
\bauthor{\bsnm{{Kunze}}, \binits{M.}},
\bauthor{\bsnm{{Langematz}}, \binits{U.}},
\bauthor{\bsnm{{Marsh}}, \binits{D.R.}},
\bauthor{\bsnm{{Maycock}}, \binits{A.C.}},
\bauthor{\bsnm{{Misios}}, \binits{S.}},
\bauthor{\bsnm{{Rodger}}, \binits{C.J.}},
\bauthor{\bsnm{{Scaife}}, \binits{A.A.}},
\bauthor{\bsnm{{Sepp{\"a}l{\"a}}}, \binits{A.}},
\bauthor{\bsnm{{Shangguan}}, \binits{M.}},
\bauthor{\bsnm{{Sinnhuber}}, \binits{M.}},
\bauthor{\bsnm{{Tourpali}}, \binits{K.}},
\bauthor{\bsnm{{Usoskin}}, \binits{I.}},
\bauthor{\bsnm{{van de Kamp}}, \binits{M.}},
\bauthor{\bsnm{{Verronen}}, \binits{P.T.}},
\bauthor{\bsnm{{Versick}}, \binits{S.}}:
\byear{2017},
\batitle{{Solar forcing for CMIP6 (v3.2)}}.
\bjtitle{Geoscientific Model Development}
\bvolume{10},
\bfpage{2247}.
\doiurl{https://doi.org/10.5194/gmd-10-2247-2017}.
\adsurl{2017GMD....10.2247M}.
\end{barticle}
\endbibitem

\bibitem[\protect\citeauthoryear{{McClintock}, {Snow}, and
  {Woods}}{2005}]{McClintock2005}
\begin{barticle}
\bauthor{\bsnm{{McClintock}}, \binits{W.E.}},
\bauthor{\bsnm{{Snow}}, \binits{M.}},
\bauthor{\bsnm{{Woods}}, \binits{T.N.}}:
\byear{2005},
\batitle{{Solar-Stellar Irradiance Comparison Experiment II (SOLSTICE II):
  Pre-Launch and On-Orbit Calibrations}}.
\bjtitle{\solphys}
\bvolume{230},
\bfpage{259}.
\doiurl{https://doi.org/10.1007/s11207-005-1585-5}.
\adsurl{2005SoPh..230..259M}.
\end{barticle}
\endbibitem

\bibitem[\protect\citeauthoryear{{Muscheler} et~al.}{2007}]{muscheler2007}
\begin{barticle}
\bauthor{\bsnm{{Muscheler}}, \binits{R.}},
\bauthor{\bsnm{{Joos}}, \binits{F.}},
\bauthor{\bsnm{{Beer}}, \binits{J.}},
\bauthor{\bsnm{{M{\"u}ller}}, \binits{S.A.}},
\bauthor{\bsnm{{Vonmoos}}, \binits{M.}},
\bauthor{\bsnm{{Snowball}}, \binits{I.}}:
\byear{2007},
\batitle{{Solar activity during the last 1000 yr inferred from radionuclide
  records}}.
\bjtitle{Quaternary Science Reviews}
\bvolume{26},
\bfpage{82}.
\doiurl{https://doi.org/10.1016/j.quascirev.2006.07.012}.
\adsurl{2007QSRv...26...82M}.
\end{barticle}
\endbibitem

\bibitem[\protect\citeauthoryear{{Namekata} et~al.}{2023}]{Namekata2023}
\begin{barticle}
\bauthor{\bsnm{{Namekata}}, \binits{K.}},
\bauthor{\bsnm{{Toriumi}}, \binits{S.}},
\bauthor{\bsnm{{Airapetian}}, \binits{V.S.}},
\bauthor{\bsnm{{Shoda}}, \binits{M.}},
\bauthor{\bsnm{{Watanabe}}, \binits{K.}},
\bauthor{\bsnm{{Notsu}}, \binits{Y.}}:
\byear{2023},
\batitle{{Reconstructing the XUV Spectra of Active Sun-like Stars Using Solar
  Scaling Relations with Magnetic Flux}}.
\bjtitle{\apj}
\bvolume{945},
\bfpage{147}.
\doiurl{https://doi.org/10.3847/1538-4357/acbe38}.
\adsurl{2023ApJ...945..147N}.
\end{barticle}
\endbibitem

\bibitem[\protect\citeauthoryear{{Neupert}}{1965}]{Neupert1965}
\begin{barticle}
\bauthor{\bsnm{{Neupert}}, \binits{W.M.}}:
\byear{1965},
\batitle{{Intensity variations in the solar extreme ultraviolet spectrum
  observed by OSO-1}}.
\bjtitle{Annales d'Astrophysique}
\bvolume{28},
\bfpage{446}.
\adsurl{1965AnAp...28..446N}.
\end{barticle}
\endbibitem

\bibitem[\protect\citeauthoryear{{Owens} et~al.}{2017}]{owens2017}
\begin{barticle}
\bauthor{\bsnm{{Owens}}, \binits{M.J.}},
\bauthor{\bsnm{{Lockwood}}, \binits{M.}},
\bauthor{\bsnm{{Riley}}, \binits{P.}},
\bauthor{\bsnm{{Linker}}, \binits{J..}}:
\byear{2017},
\batitle{{Sunward Strahl: A Method to Unambiguously Determine Open Solar Flux
  from In Situ Spacecraft Measurements Using Suprathermal Electron Data}}.
\bjtitle{Journal of Geophysical Research: Space Physics}
\bvolume{122},
\bfpage{980}.
\doiurl{https://doi.org/10.1002/2017JA024631}.
\end{barticle}
\endbibitem

\bibitem[\protect\citeauthoryear{{Penza} et~al.}{2022}]{penza2022}
\begin{barticle}
\bauthor{\bsnm{{Penza}}, \binits{V.}},
\bauthor{\bsnm{{Berrilli}}, \binits{F.}},
\bauthor{\bsnm{L.}, \binits{B.}},
\bauthor{\bsnm{{Cantoresi}}, \binits{M.}},
\bauthor{\bsnm{{Criscuoli}}, \binits{S.}},
\bauthor{\bsnm{{Giobbi}}, \binits{P.}}:
\byear{2022},
\batitle{{Total Solar Irradiance during the Last Five Centuries}}.
\bjtitle{\apjl}
\bvolume{937},
\bfpage{84}.
\doiurl{https://doi.org/10.3847/1538-4357/ac8a4b}.
\end{barticle}
\endbibitem

\bibitem[\protect\citeauthoryear{{Penza} et~al.}{2024}]{penza2024}
\begin{barticle}
\bauthor{\bsnm{{Penza}}, \binits{V.}},
\bauthor{\bsnm{{Bertello}}, \binits{L.}},
\bauthor{\bsnm{{Cantoresi}}, \binits{M.}},
\bauthor{\bsnm{{Criscuoli}}, \binits{S.}},
\bauthor{\bsnm{{Lucaferri}}, \binits{L.}},
\bauthor{\bsnm{{Reda}}, \binits{R.}},
\bauthor{\bsnm{{Ulzega}}, \binits{S.}},
\bauthor{\bsnm{{Berrilli}}, \binits{F.}}:
\byear{2024},
\batitle{{Reconstruction of the Total Solar Irradiance During the Last
  Millennium}}.
\bjtitle{\apj}
\bvolume{976},
\bfpage{11}.
\doiurl{https://doi.org/10.3847/1538-4357/ad7c49}.
\adsurl{2024ApJ...976...11P}.
\end{barticle}
\endbibitem

\bibitem[\protect\citeauthoryear{{Petrie}, {Criscuoli}, and
  {Bertello}}{2021}]{Petrie2021}
\begin{bchapter}
\bauthor{\bsnm{{Petrie}}, \binits{G.}},
\bauthor{\bsnm{{Criscuoli}}, \binits{S.}},
\bauthor{\bsnm{{Bertello}}, \binits{L.}}:
\byear{2021},
\bctitle{{Solar Magnetism and Radiation}}.
In: \beditor{\bsnm{{Raouafi}}, \binits{N.E.}},
\beditor{\bsnm{{Vourlidas}}, \binits{A.}} (eds.)
\bbtitle{Solar Physics and Solar Wind}
\bseriesno{1},
\bfpage{83}.
\doiurl{https://doi.org/10.1002/9781119815600.ch3}.
\adsurl{2021GMS...258...83P}.
\end{bchapter}
\endbibitem

\bibitem[\protect\citeauthoryear{{Pienitz} and {Vincent}}{2000}]{Pienitz2000}
\begin{barticle}
\bauthor{\bsnm{{Pienitz}}, \binits{R.}},
\bauthor{\bsnm{{Vincent}}, \binits{W.F.}}:
\byear{2000},
\batitle{{Effect of climate change relative to ozone depletion on UV exposure
  in subarctic lakes}}.
\bjtitle{\nat}
\bvolume{404},
\bfpage{484}.
\doiurl{https://doi.org/10.1038/35006616}.
\adsurl{2000Natur.404..484P}.
\end{barticle}
\endbibitem

\bibitem[\protect\citeauthoryear{{Reda} et~al.}{2022}]{Reda2022}
\begin{barticle}
\bauthor{\bsnm{{Reda}}, \binits{R.}},
\bauthor{\bsnm{{Di Mauro}}, \binits{M.P.}},
\bauthor{\bsnm{{Giovannelli}}, \binits{L.}},
\bauthor{\bsnm{{Alberti}}, \binits{T.}},
\bauthor{\bsnm{{Berrilli}}, \binits{F.}},
\bauthor{\bsnm{{Corsaro}}, \binits{E.}}:
\byear{2022},
\batitle{{A Synergic Strategy to Characterize the Habitability Conditions of
  Exoplanets Hosted by Solar-Type Stars}}.
\bjtitle{Frontiers in Astronomy and Space Sciences}
\bvolume{9},
\bfpage{909268}.
\doiurl{https://doi.org/10.3389/fspas.2022.909268}.
\adsurl{2022FrASS...9.9268R}.
\end{barticle}
\endbibitem

\bibitem[\protect\citeauthoryear{{Reda} et~al.}{2023}]{Reda2023}
\begin{barticle}
\bauthor{\bsnm{{Reda}}, \binits{R.}},
\bauthor{\bsnm{{Giovannelli}}, \binits{L.}},
\bauthor{\bsnm{{Alberti}}, \binits{T.}},
\bauthor{\bsnm{{Berrilli}}, \binits{F.}},
\bauthor{\bsnm{{Bertello}}, \binits{L.}},
\bauthor{\bsnm{{Del Moro}}, \binits{D.}},
\bauthor{\bsnm{{Di Mauro}}, \binits{M.P.}},
\bauthor{\bsnm{{Giobbi}}, \binits{P.}},
\bauthor{\bsnm{{Penza}}, \binits{V.}}:
\byear{2023},
\batitle{{The exoplanetary magnetosphere extension in Sun-like stars based on
  the solar wind-solar UV relation}}.
\bjtitle{\mnras}
\bvolume{519},
\bfpage{6088}.
\doiurl{https://doi.org/10.1093/mnras/stac3825}.
\adsurl{2023MNRAS.519.6088R}.
\end{barticle}
\endbibitem

\bibitem[\protect\citeauthoryear{{Richard} et~al.}{2024}]{Richard2024}
\begin{barticle}
\bauthor{\bsnm{{Richard}}, \binits{E.}},
\bauthor{\bsnm{{Coddington}}, \binits{O.}},
\bauthor{\bsnm{{Harber}}, \binits{D.}},
\bauthor{\bsnm{{Chambliss}}, \binits{M.}},
\bauthor{\bsnm{{Penton}}, \binits{S.}},
\bauthor{\bsnm{{Brooks}}, \binits{K.}},
\bauthor{\bsnm{{Charbonneau}}, \binits{L.}},
\bauthor{\bsnm{{Peck}}, \binits{C.}},
\bauthor{\bsnm{{B{\'e}land}}, \binits{S.}},
\bauthor{\bsnm{{Pilewskie}}, \binits{P.}},
\bauthor{\bsnm{{Woods}}, \binits{T.}}:
\byear{2024},
\batitle{{Advancements in solar spectral irradiance measurements by the TSIS-1
  spectral irradiance monitor and its role for long-term data continuity}}.
\bjtitle{Journal of Space Weather and Space Climate}
\bvolume{14},
\bfpage{10}.
\doiurl{https://doi.org/10.1051/swsc/2024008}.
\adsurl{2024JSWSC..14...10R}.
\end{barticle}
\endbibitem

\bibitem[\protect\citeauthoryear{{Schindhelm} et~al.}{2015}]{Schindelm15}
\begin{barticle}
\bauthor{\bsnm{{Schindhelm}}, \binits{R.}},
\bauthor{\bsnm{{Stern}}, \binits{S.A.}},
\bauthor{\bsnm{{Gladstone}}, \binits{R.}},
\bauthor{\bsnm{{Zangari}}, \binits{A.}}:
\byear{2015},
\batitle{{Pluto and Charon's UV spectra from IUE to New Horizons}}.
\bjtitle{Icarus}
\bvolume{246},
\bfpage{206}.
\doiurl{https://doi.org/10.1016/j.icarus.2014.03.003}.
\adsurl{2015Icar..246..206S}.
\end{barticle}
\endbibitem

\bibitem[\protect\citeauthoryear{{Segura} et~al.}{2005}]{Segura2005}
\begin{barticle}
\bauthor{\bsnm{{Segura}}, \binits{A.}},
\bauthor{\bsnm{{Kasting}}, \binits{J.F.}},
\bauthor{\bsnm{{Meadows}}, \binits{V.}},
\bauthor{\bsnm{{Cohen}}, \binits{M.}},
\bauthor{\bsnm{{Scalo}}, \binits{J.}},
\bauthor{\bsnm{{Crisp}}, \binits{D.}},
\bauthor{\bsnm{{Butler}}, \binits{R.A.H.}},
\bauthor{\bsnm{{Tinetti}}, \binits{G.}}:
\byear{2005},
\batitle{{Biosignatures from Earth-Like Planets Around M Dwarfs}}.
\bjtitle{Astrobiology}
\bvolume{5},
\bfpage{706}.
\doiurl{https://doi.org/10.1089/ast.2005.5.706}.
\adsurl{2005AsBio...5..706S}.
\end{barticle}
\endbibitem

\bibitem[\protect\citeauthoryear{{Solanki}, {Krivova}, and
  {Haigh}}{2013}]{solanki2013}
\begin{barticle}
\bauthor{\bsnm{{Solanki}}, \binits{S.K.}},
\bauthor{\bsnm{{Krivova}}, \binits{N.A.}},
\bauthor{\bsnm{{Haigh}}, \binits{J.D.}}:
\byear{2013},
\batitle{{Solar Irradiance Variability and Climate}}.
\bjtitle{Annual Review of Astronomy and Astrophysics}
\bvolume{51},
\bfpage{311}.
\doiurl{https://doi.org/10.1146/annurev-astro-082812-141007}.
\end{barticle}
\endbibitem

\bibitem[\protect\citeauthoryear{{Sowmya} et~al.}{2025}]{Sowmya2025}
\begin{barticle}
\bauthor{\bsnm{{Sowmya}}, \binits{K.}},
\bauthor{\bsnm{{Snow}}, \binits{M.}},
\bauthor{\bsnm{{Shapiro}}, \binits{A.I.}},
\bauthor{\bsnm{{Krivova}}, \binits{N.A.}},
\bauthor{\bsnm{{Chatzistergos}}, \binits{T.}},
\bauthor{\bsnm{{Solanki}}, \binits{S.K.}}:
\byear{2025},
\batitle{{Solar Variability in the Mg II h and k Lines}}.
\bjtitle{\apj}
\bvolume{980},
\bfpage{173}.
\doiurl{https://doi.org/10.3847/1538-4357/ada6af}.
\adsurl{2025ApJ...980..173S}.
\end{barticle}
\endbibitem

\bibitem[\protect\citeauthoryear{{Spinelli} et~al.}{2023}]{Spinelli2023}
\begin{barticle}
\bauthor{\bsnm{{Spinelli}}, \binits{R.}},
\bauthor{\bsnm{{Borsa}}, \binits{F.}},
\bauthor{\bsnm{{Ghirlanda}}, \binits{G.}},
\bauthor{\bsnm{{Ghisellini}}, \binits{G.}},
\bauthor{\bsnm{{Haardt}}, \binits{F.}}:
\byear{2023},
\batitle{{The ultraviolet habitable zone of exoplanets}}.
\bjtitle{\mnras}
\bvolume{522},
\bfpage{1411}.
\doiurl{https://doi.org/10.1093/mnras/stad928}.
\adsurl{2023MNRAS.522.1411S}.
\end{barticle}
\endbibitem

\bibitem[\protect\citeauthoryear{{Steinegger}, {Brandt}, and
  {Haupt}}{1996}]{Steinegger1996}
\begin{barticle}
\bauthor{\bsnm{{Steinegger}}, \binits{M.}},
\bauthor{\bsnm{{Brandt}}, \binits{P.N.}},
\bauthor{\bsnm{{Haupt}}, \binits{H.F.}}:
\byear{1996},
\batitle{{Sunspot irradiance deficit, facular excess, and the energy balance of
  solar active regions.}}
\bjtitle{\aap}
\bvolume{310},
\bfpage{635}.
\adsurl{1996A&A...310..635S}.
\end{barticle}
\endbibitem

\bibitem[\protect\citeauthoryear{{Stuiver} and {Braziunas}}{1998}]{Stuiver1998}
\begin{barticle}
\bauthor{\bsnm{{Stuiver}}, \binits{M.}},
\bauthor{\bsnm{{Braziunas}}, \binits{T.F.}}:
\byear{1998},
\batitle{{Anthropogenic and solar components of hemispheric $^{14}$C}}.
\bjtitle{\grl}
\bvolume{25},
\bfpage{329}.
\doiurl{https://doi.org/10.1029/97GL03694}.
\adsurl{1998GeoRL..25..329S}.
\end{barticle}
\endbibitem

\bibitem[\protect\citeauthoryear{{Usoskin}}{2017}]{usoskin2017}
\begin{barticle}
\bauthor{\bsnm{{Usoskin}}, \binits{I.G.}}:
\byear{2017},
\batitle{{A history of solar activity over millennia}}.
\bjtitle{Living Reviews in Solar Physics}
\bvolume{14},
\bfpage{3}.
\doiurl{https://doi.org/10.1007/s41116-017-0006-9}.
\adsurl{2017LRSP...14....3U}.
\end{barticle}
\endbibitem

\bibitem[\protect\citeauthoryear{{Usoskin} et~al.}{2016}]{usoskin2016}
\begin{barticle}
\bauthor{\bsnm{{Usoskin}}, \binits{I.G.}},
\bauthor{\bsnm{{Gallet}}, \binits{Y.}},
\bauthor{\bsnm{{Lopes}}, \binits{F.}},
\bauthor{\bsnm{{Kovaltsov}}, \binits{G.A.}},
\bauthor{\bsnm{{Hulot}}, \binits{G.}}:
\byear{2016},
\batitle{{Solar activity during the Holocene: the Hallstatt cycle and its
  consequence for grand minima and maxima}}.
\bjtitle{\AA}
\bvolume{587},
\bfpage{A150}.
\doiurl{https://doi.org/10.1051/0004-6361/201527295}.
\adsurl{https://www.aanda.org/articles/aa/pdf/2016/03/aa27295-15.pdf}.
\end{barticle}
\endbibitem

\bibitem[\protect\citeauthoryear{{Usoskin} et~al.}{2021}]{usoskin2021}
\begin{barticle}
\bauthor{\bsnm{{Usoskin}}, \binits{I.G.}},
\bauthor{\bsnm{{Solanki}}, \binits{S.K.}},
\bauthor{\bsnm{{Krivova}}, \binits{N.A.}},
\bauthor{\bsnm{{Hofer}}, \binits{B.}},
\bauthor{\bsnm{G.A.}, \binits{K.}},
\bauthor{\bsnm{{Wacker}}, \binits{L.}},
\bauthor{\bsnm{{Brehm}}, \binits{N.}},
\bauthor{\bsnm{{Kromer}}, \binits{B.}}:
\byear{2021},
\batitle{{Solar cyclic activity over the last millennium reconstructed from
  annual 14C data}}.
\bjtitle{\aap}
\bvolume{649}.
\doiurl{https://doi.org/10.1051/0004-6361/202140711}.
\adsurl{2017LRSP...14....3U}.
\end{barticle}
\endbibitem

\bibitem[\protect\citeauthoryear{{Vecchio} et~al.}{2017}]{Vecchio2017}
\begin{barticle}
\bauthor{\bsnm{{Vecchio}}, \binits{A.}},
\bauthor{\bsnm{{Lepreti}}, \binits{F.}},
\bauthor{\bsnm{{Laurenza}}, \binits{M.}},
\bauthor{\bsnm{{Alberti}}, \binits{T.}},
\bauthor{\bsnm{{Carbone}}, \binits{V.}}:
\byear{2017},
\batitle{{Connection between solar activity cycles and grand minima
  generation}}.
\bjtitle{\aap}
\bvolume{599},
\bfpage{A58}.
\doiurl{https://doi.org/10.1051/0004-6361/201629758}.
\adsurl{2017A&A...599A..58V}.
\end{barticle}
\endbibitem

\bibitem[\protect\citeauthoryear{{Volobuev}}{2009}]{Volobuev}
\begin{barticle}
\bauthor{\bsnm{{Volobuev}}, \binits{D.M.}}:
\byear{2009},
\batitle{{The Shape of The Sunspot Cycle: A One-Parameter Fit}}.
\bjtitle{\solphys}
\bvolume{258},
\bfpage{319}.
\doiurl{https://doi.org/1 10.1007/s11207-009-9429-3}.
\end{barticle}
\endbibitem

\bibitem[\protect\citeauthoryear{{Vonmoos}, {Beer}, and
  {Muscheler}}{2006}]{Vonmoos2006}
\begin{barticle}
\bauthor{\bsnm{{Vonmoos}}, \binits{M.}},
\bauthor{\bsnm{{Beer}}, \binits{J.}},
\bauthor{\bsnm{{Muscheler}}, \binits{R.}}:
\byear{2006},
\batitle{{Large variations in Holocene solar activity: Constraints from
  $^{10}$Be in the Greenland Ice Core Project ice core}}.
\bjtitle{Journal of Geophysical Research (Space Physics)}
\bvolume{111},
\bfpage{A10105}.
\doiurl{https://doi.org/10.1029/2005JA011500}.
\adsurl{2006JGRA..11110105V}.
\end{barticle}
\endbibitem

\bibitem[\protect\citeauthoryear{{Wenzler} et~al.}{2006}]{wenzler}
\begin{barticle}
\bauthor{\bsnm{{Wenzler}}, \binits{T.}},
\bauthor{\bsnm{{Solanki}}, \binits{S.K.}},
\bauthor{\bsnm{{Krivova}}, \binits{N.A.}},
\bauthor{\bsnm{{Fröhlich}}, \binits{C.}}:
\byear{2006},
\batitle{{Reconstruction of solar irradiance variations in cycles 21–23 based
  on surface magnetic fields}}.
\bjtitle{\aap}
\bvolume{460},
\bfpage{583}.
\doiurl{https://doi.org/10.1051/0004-6361:20065752}.
\end{barticle}
\endbibitem

\bibitem[\protect\citeauthoryear{{Woods} and {DeLand}}{2021}]{Woods21}
\begin{barticle}
\bauthor{\bsnm{{Woods}}, \binits{T.N.}},
\bauthor{\bsnm{{DeLand}}, \binits{M.T.}}:
\byear{2021},
\batitle{{An Improved Solar Spectral Irradiance Composite Record}}.
\bjtitle{Earth and Space Science}
\bvolume{8},
\bfpage{e01740}.
\doiurl{https://doi.org/10.1029/2021EA001740}.
\adsurl{2021E&SS....801740W}.
\end{barticle}
\endbibitem

\end{thebibliography}


\end{document}